\documentclass[aps,prb,twocolumn,showpacs, notitlepage]{revtex4-2}
\usepackage{graphicx}
\usepackage{natbib}
\usepackage[utf8]{inputenc}
\usepackage{epstopdf}
\setcitestyle{super}
\usepackage{color}
\usepackage{amsmath}
\usepackage[normalem]{ulem}
\usepackage{pdfpages}
\usepackage{hyperref}
\makeatletter
\AtBeginDocument{\let\LS@rot\@undefined}
\makeatother

\newcommand*{\citen}[1]{%
  \begingroup
    \romannumeral-`\x 
    \setcitestyle{numbers}%
    \cite{#1}%
  \endgroup   
}

\usepackage{hyphenat}
\begin{document}

\preprint{APS/123-QED}

\title{Pressure-induced ferromagnetism in the topological semimetal EuCd$_2$As$_2$}
\author{Elena Gati$^{1,2,3}$}\email{elena.gati@cpfs.mpg.de}
\author{Sergey L. Bud'ko$^{1,2}$}
\author{Lin-Lin Wang$^{1}$}
\author{Adrian Valadkhani$^4$}
\author{Ritu Gupta$^5$}
\author{Brinda Kuthanazhi$^{1,2}$}
\author{Li Xiang$^{1,2}$}
\author{John M. Wilde$^{1,2}$}
\author{Aashish Sapkota$^{1,2}$}
\author{Zurab Guguchia$^5$}
\author{Rustem Khasanov$^{5}$}
\author{Roser Valent\'{i}$^4$}
\author{Paul C. Canfield$^{1,2}$}\email{canfield@ameslab.gov}

\address{$^{1}$ Ames Laboratory, US Department of Energy, Iowa State University, Ames, Iowa 50011, USA}
\address{$^{2}$ Department of Physics and Astronomy, Iowa State University, Ames, Iowa 50011, USA}
\address{$^{3}$ Max Planck Institute for Chemical Physics of Solids, 01187 Dresden, Germany}
\address{$^{4}$ Institute for Theoretical Physics, Goethe-Universit\"{a}t Frankfurt, Max-von-Laue-Strasse 1, 60438 Frankfurt am Main, Germany}
\address{$^{5}$ Laboratory for Muon Spin Spectroscopy, Paul Scherrer Institute, 5232 Villigen PSI, Switzerland}

\date{\today}

\begin{abstract}
The antiferromagnet and semimetal EuCd$_2$As$_2$ has recently attracted a lot of attention due to a wealth of topological phases arising
from the interplay of topology and magnetism.
In particular, the presence of a single pair of Weyl points
is predicted for a ferromagnetic configuration of Eu spins
along the $c$-axis in EuCd$_2$As$_2$. In the search for such phases, we investigate here the effects of hydrostatic pressure in EuCd$_2$As$_2$. For that, we present specific heat, transport and $\mu$SR measurements under hydrostatic pressure up to $\sim\,2.5\,$GPa, combined with 
{\it ab initio} density functional theory (DFT) calculations. Experimentally, we establish that the ground state of EuCd$_2$As$_2$ changes from in-plane antiferromagnetic (AFM$_{ab}$) to ferromagnetic at a critical pressure of $\,\approx\,$2\,GPa, which is likely characterized by the moments dominantly lying within the $ab$ plane (FM$_{ab}$). The AFM$_{ab}$-FM$_{ab}$ transition at such a relatively low pressure is supported by our DFT calculations. Furthermore, our experimental and theoretical results indicate that EuCd$_2$As$_2$ moves closer to the sought-for FM$_c$ state (moments $\parallel$ $c$) with increasing pressure further. We predict that a pressure of $\approx$\,23\,GPa will stabilize the FM$_c$ state, if Eu remains in a 2+ valence state. Thus, our work establishes hydrostatic pressure as a key tuning parameter that (i) allows for a continuous tuning between magnetic ground states in a single sample of EuCd$_2$As$_2$ and (ii) enables the exploration of the interplay between magnetism and topology and thereby motivates a series of future experiments on this magnetic Weyl semimetal.
\end{abstract}

\pacs{xxx}

\maketitle

\section{Introduction}

Topological materials have been at the center of many research 
activities in recent years due to the presence of a plethora of 
exotic phenomena, of relevance  not only 
from a fundamental perspective \cite{Wan11} but for applications \cite{Hasan10,Yan17,Burkov16,Zhang19,Smejkal17} as well. In this context, Weyl semimetals \cite{Yan17,Armitage18,Burkov16} have emerged as particularly multifaceted realizations of topological
materials displaying anomalous transport phenomena, including the anomalous Hall effect, \cite{Burkov11,Xu11} negative longitudinal magnetoresistance \cite{Son13} and non-local transport \cite{Parameswaran14}.

A Weyl semimetal can be created from a Dirac semimetal by breaking either inversion symmetry \cite{Liu14} or time reversal symmetry \cite{Yan17,Armitage18,Burkov16,heinsdorf2021}. Thus, Weyl nodes can only be found either in non-centrosymmetric crystals or in magnetic materials, providing a seemingly straight-forward path to identify new Weyl semimetals. However, in real materials, the observation of the intriguing features of Weyl physics is often complicated \cite{Chang18,Arnold16} by (i) additional non-topological Fermi surface pockets and (ii) the occurrence of multiple Weyl nodes. To overcome the latter issue, research is recently focussing on magnetic systems, since the breaking of time-reversal symmetry allows, in principle, the realization of a single pair of Weyl nodes \cite{Armitage18}, which, in the presence of inversion symmetry, have to be located at the same energy. In contrast, when the crystal structure breaks inversion symmetry, the minimal number of Weyl nodes is 4, although often is found to be higher, such as 24 in the TaAs structural family of compounds \cite{Yan17,Weng15,Huang15b}. 


Whereas several magnetic Weyl semimetals were proposed and studied experimentally \cite{Borisenko19,Liu18,Ye18,Kang20,Guguchia20,Destraz20}, there is a continuing interest in identifying new candidate materials, which display unambiguous signatures of Weyl physics, free from the above-mentioned complexities. Surveys of space groups \cite{Hua18} and band structure calculations \cite{Wang19} have conditionally identified EuCd$_2$As$_2$ (centrosymmetric space group $P\bar{3}m$1 \cite{Artmann96}) as a candidate for 
magnetic Weyl semimetal if it orders ferromagnetically (FM) with the moments parallel to the $c$ axis. In such a FM state, EuCd$_2$As$_2$ would show the minimal number of Weyl nodes in close proximity to the Fermi level without any further manipulation, such as the application of magnetic field. 

However, experimental studies indicate that  the Eu spins in EuCd$_2$As$_2$ order antiferromagnetically (AFM) at ambient pressure below $T_\textrm{N}\,\simeq\,9\,$K \cite{Wang16,Rahn18}. Although controversially discussed, most papers refer to the AFM structure as the A-type, i.e., FM layers which stack in an AFM fashion along the $c$ axis, with the moment direction confined in the $ab$ plane \cite{Rahn18,Soh19}. Albeit this order breaks the in-plane three-fold symmetry and thus hinders the formation of Weyl nodes, there are several experimental observations that indicate that Weyl physics can manifest itself in EuCd$_2$As$_2$ due to the proximity to ferromagnetism. Among those are, e.g., the report of a single pair of Weyl nodes in EuCd$_2$As$_2$ in the $c$ polarized state that is stabilized by very moderate magnetic fields \cite{Soh19}, as well as the claim of Weyl nodes in the paramagnetic state whose presence was attributed to quasi-static and quasi-long range ferromagnetic fluctuations \cite{Ma19}. Overall, these results underline that EuCd$_2$As$_2$ is a promising host for investigations of the interplay of topology and magnetism \cite{Niu19}. This notion has been further supported by transport measurements at temperatures around the antiferromagnetic transition \cite{Xu21}. \\
Actually, it was recently shown that small differences in the synthesis procedure of this compound can result in samples of EuCd$_2$As$_2$ with FM order and moments lying in the $ab$ plane \cite{Jo20,Sanjeewa20,Taddei20}. Whereas this explicitly demonstrates that a FM state in EuCd$_2$As$_2$ is very close (in some compositional space) to the AFM one, and suggests a high degree of tunability of the magnetic properties of EuCd$_2$As$_2$, this also can cause severe complications in comparing published results on different crystals grown via different routes. \\
Here, we report the effect of hydrostatic pressure, $p$, up to 2.5\,GPa on the magnetic properties of AFM EuCd$_2$As$_2$ by presenting specific heat, transport and $\mu$SR measurements, combined with DFT calculations of total energies of different magnetic configurations up to 25\,GPa. Our experiments unambiguously identify a transition to an order with a pronounced FM component, which we refer to as FM order hereafter, at $p_c\,\approx\,2$\,GPa. This FM order is likely characterized by an in-plane orientation of the magnetic moments, as supported by our DFT calculations. Further, our calculations predict a change of the moment orientation at $\approx\,$23\,GPa to an out-of-plane configuration, which is the required ground state for realizing a single pair of Weyl nodes in this material. Our results therefore clearly identify pressure as an experimental tuning parameter that allows for studies of the correlation of the topological properties of EuCd$_2$As$_2$ with a change of its magnetic ground state in a single sample.

The paper is organized as follows. Section~\ref{sec:methods} summarizes the experimental settings and theoretical methods used in this work. In Section \ref{sec:Results}, we discuss our results, starting with a discussion of experimental phase diagram (Sec.\,\ref{sec:phasediagram}, followed by a presentation of our $\mu$SR data (Sec.\,\ref{sec:microscopic-magnetism}) and anisotropic magnetoresistance data (Sec.\,\ref{sec:anisotropic-magnetoresistance}) and concluded by a discussion of our results from DFT calculations (Sec.\,\ref{sec:DFTcalculations}). Finally in Section \ref{sec:conclusions}, we present our conclusions and an outlook. 

\section{Methods}\label{sec:methods}

Single crystals of EuCd$_2$As$_2$ were grown out of Sn flux with the following procedure; The elements with an initial stoichiometry of Eu:Cd:As:Sn = 1:2:2:10 were put into a fritted alumina crucible \cite{Canfield16} (sold by LSP Ceramics as a Canfield Crucible Set \cite{CanfieldCCS}) and sealed in fused silica tube under a partial pressure of argon. The thus prepared ampoule was heated up to 900$^\circ$C over 24 hours, and held there for 20 hours. This was followed by a slow cooling to 550$^\circ$C over 200 hours, and decanting of the excess flux using a centrifuge \cite{Canfield19}. The crystals were characterized by the means of powder x-ray diffraction as well as magnetic measurements. The latter measurements confirmed that these crystals undergo an AFM transition\cite{Jo20} at $T_\textrm{N}\,\sim\,9.5$\,K.

Specific heat under pressure was measured on a single crystal using the AC calorimetry technique, as described in detail in Ref. \citen{Gati19}. Resistance under pressure was measured in a four-point configuration. Contacts were made using Epo-tek H20E silver epoxy. Unfortunately, despite using different contact materials and routes of surface preparation, it was not possible to get two-point resistances smaller than several tens of Ohms. Given that we are mostly interested in tracking anomalies in the temperature- and field dependence of $R(T)$, which we carefully  measure by reducing the measurement current, we refrain from correcting the presented transport data for geometrical factors. For specific heat as well as transport measurements, the cryogenic environment was provided by a Quantum Design Physical Property Measurement System. Pressure was generated in a piston-cylinder double-wall pressure cell with the outer cylinder made out of CuBe and the inner cylinder out of Ni-Cr-Al alloy (see Ref.\,\citen{Budko84} for a very similar design). A mixture of 4:6 light mineral oil:n-pentane was used as a pressure-transmitting medium. This medium solidifies at $p\,\approx\,3-4\,$GPa at room temperature\cite{Torikachvili15}, thus ensuring hydrostatic pressure application over the available pressure range. Pressure at low temperatures was determined from the shift of the superconducting transition temperature of elemental lead (Pb)\cite{Eiling81}.

$\mu$SR measurements under pressure were performed in a $^3$He cryostat at the $\mu$E1 beamline at the Paul-Scherrer-Institute in Villigen, Switzerland, by using the GPD spectrometer. Typically, $5 - 10\,\cdot\,10^6$ positron events were counted for each data point. A large number of single crystals of total mass of $\approx\,2\,$g were placed inside a pressure cell with arbitrary orientations. Both, the inner and the outer cylinder of the pressure cell are made out of MP35N alloy\cite{Khasanov16}. Daphne 7373 oil was used as a pressure-transmitting medium, which solidifies at room temperature close to 2.5\,GPa\cite{Torikachvili15}. The pressure at low temperatures was determined from the shift of the superconducting transition of elemental indium\cite{Smith67}, which was also placed in the pressure cell and measured in an independent ac susceptibility experiment.

Total energies for  EuCd$_2$As$_2$ with various Eu spin configurations
were calculated in DFT \cite{Hohenberg64,Kohn65} including spin-orbit coupling (SOC) with the  Perdew–Burke–Ernzerhof (PBE) exchange-correlation functional \cite{Perdew96}. We employed a plane-wave basis set and projector augmented wave (PAW) \cite{Bloechl94,Boechl99} method as implemented in VASP \cite{Kresse93,Kresse96,Kresse9615}. To account for the half-filled strongly localized Eu 4\textit{f} orbitals, a Hubbard-like $U$ parameter of 4.4\,eV was used which  places the f-states in the region of -0.8\,eV to -1.4\,eV,  as observed experimentally in photoemission experiments \cite{Ma19}. The analysis of the magnetism as a function of pressure was done by performing equation of state (EOS) calculations of both A-type antiferromagnetic and in-plane ferromagnetic EuCd$_2$As$_2$
with the primitive hexagonal unit cell doubled along the \textit{c}-axis (Note that the energies of different moment directions in the plane were so close that differences could not be resolved within DFT).  The corresponding reciprocal lattice was sampled on  a $\Gamma$-centered Monkhorst-Pack \cite{Monkhorst76} (11$\times$11$\times$3) \textit{k}-point mesh with a Gaussian smearing of 0.05 eV.
We used a kinetic energy cutoff of 318 eV to relax the shape of the unit cell and atomic positions at selected volumes until the absolute force on each atom is below 0.01 eV/\AA. Specifically for obtaining the  crystalline magneto-anisotropy calculations, we
 increased the $k$-point mesh  to ($22\times22\times6$),
 the kinetic energy cutoff was set to 500 eV and
 the absolute force threshold was set to 0.001 eV/\AA~ per atom. In Appendix \ref{app:lattice-parameters-theory} we present a discussion on the effects that the choice of initial settings in the DFT  calculations have on the final relaxed structures.

\section{Results and Discussion}
\label{sec:Results}

\subsection{Zero-field temperature-pressure phase diagram}
\label{sec:phasediagram}

\begin{center}
\begin{figure}
\includegraphics[width=0.9\columnwidth]{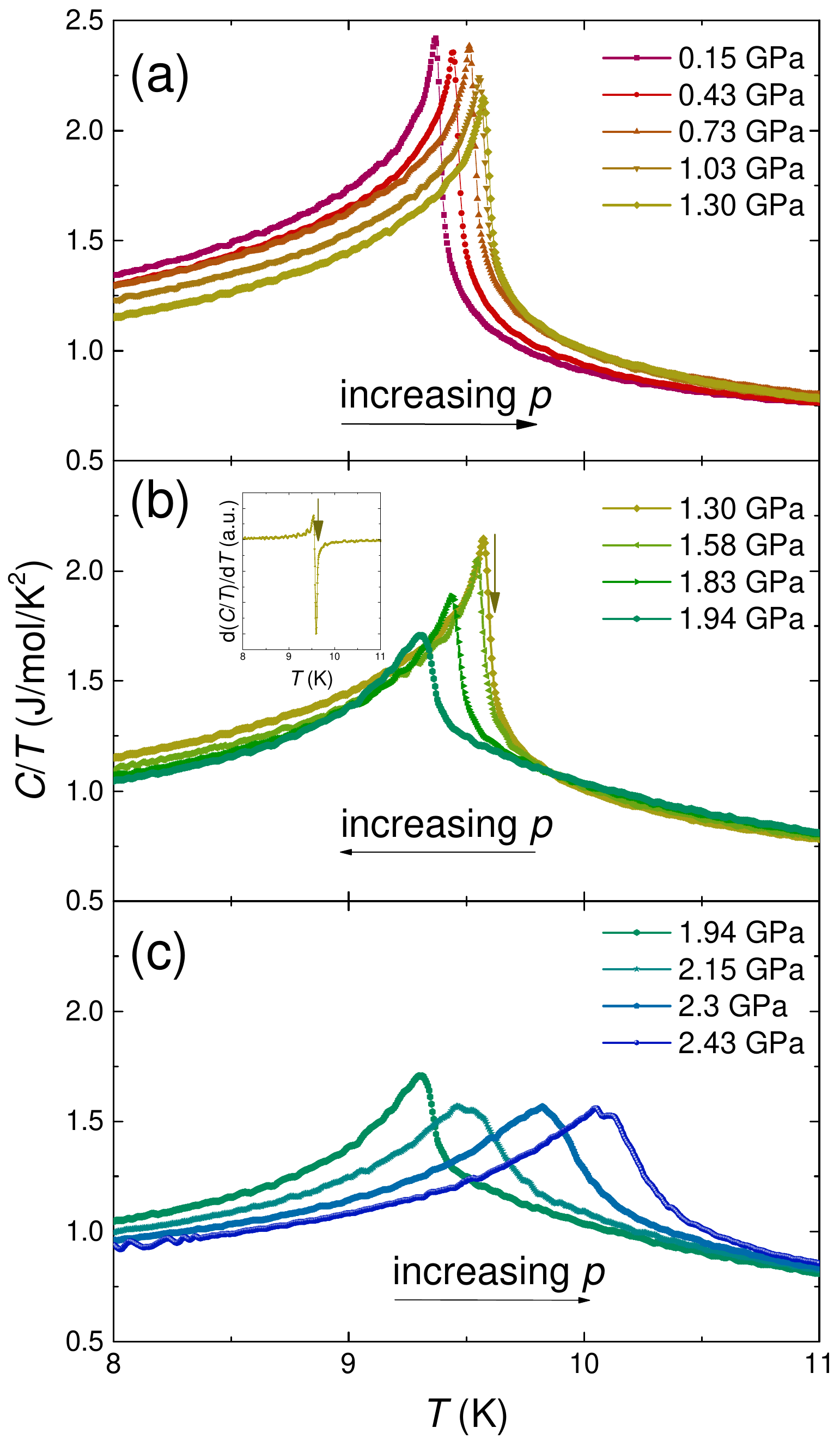} 
\caption{Selected data sets of specific heat divided by temperature, $C/T$, vs. $T$ (8\,K\,$\le\,T\,\le\,$11\,K) of EuCd$_2$As$_2$ under pressure, $p$, for $0.13\,$GPa$\,\leq\,p\,\leq\,1.30$\,GPa (a), for $1.30\,$GPa$\,\leq\,p\,\leq\,1.94$\,GPa (b) and $1.94\,$GPa$\,\leq\,p\,\leq\,2.43$\,GPa (c). The inset in (b) shows an example data set of d$(C/T)$/d$T$, the minimum of which is used to determine the transition temperature (see arrows in the inset and the main panel).}
\label{fig:specificheat}
\end{figure}
\end{center}

First, we discuss our determination of the temperature-pressure phase diagram of EuCd$_2$As$_2$ in zero field from specific heat, $C/T$, and resistance, $R$, measurements. Figure \ref{fig:specificheat} shows $C/T$ data as a function of temperature in three pressure ranges up to 2.43\,GPa. Close to ambient pressure, as represented by the 0.15\,GPa data in Fig.\,\ref{fig:specificheat}\,(a), we observe a sharp transition at $\sim\,9.4$\,K which signals the onset of antiferromagnetic order, consistent with literature \cite{Jo20}. Upon increasing pressure up to 1.30\,GPa (Fig.\,\ref{fig:specificheat}\,(a)), this feature shifts to higher temperatures, whereas almost no change can be found in the shape of the feature, i.e., in the sharpness and the maximum value of $C/T$ at the transition. Further increasing pressure beyond 1.30\,GPa (Fig.\,\ref{fig:specificheat}\,(b)) results in a clear decrease of the transition temperature up to $\,\sim\,$1.94\,GPa. Concurrently, the specific heat feature starts to broaden slightly and the maximum value decreases, implying that the associated entropy release with the magnetic ordering is distributed over a wider temperature range. At even higher pressures, up to the maximum pressure of our experiment of 2.43\,GPa (Fig.\,\ref{fig:specificheat}\,(c)), the transition temperature increases again and the increase takes place at a faster rate, compared to lower pressures. The specific heat feature is significantly broadened, indicating that fluctuations might play a role in a wider temperature range above the transition temperature. The observations described above, such as the sharp feature in the temperature-pressure phase diagram at $\approx\,$2\,GPa (see also Fig.\,\ref{fig:phasediagram} below), suggest a change of the magnetic structure of EuCd$_2$As$_2$ with hydrostatic pressure.

\begin{center}
\begin{figure}
\includegraphics[width=0.9\columnwidth]{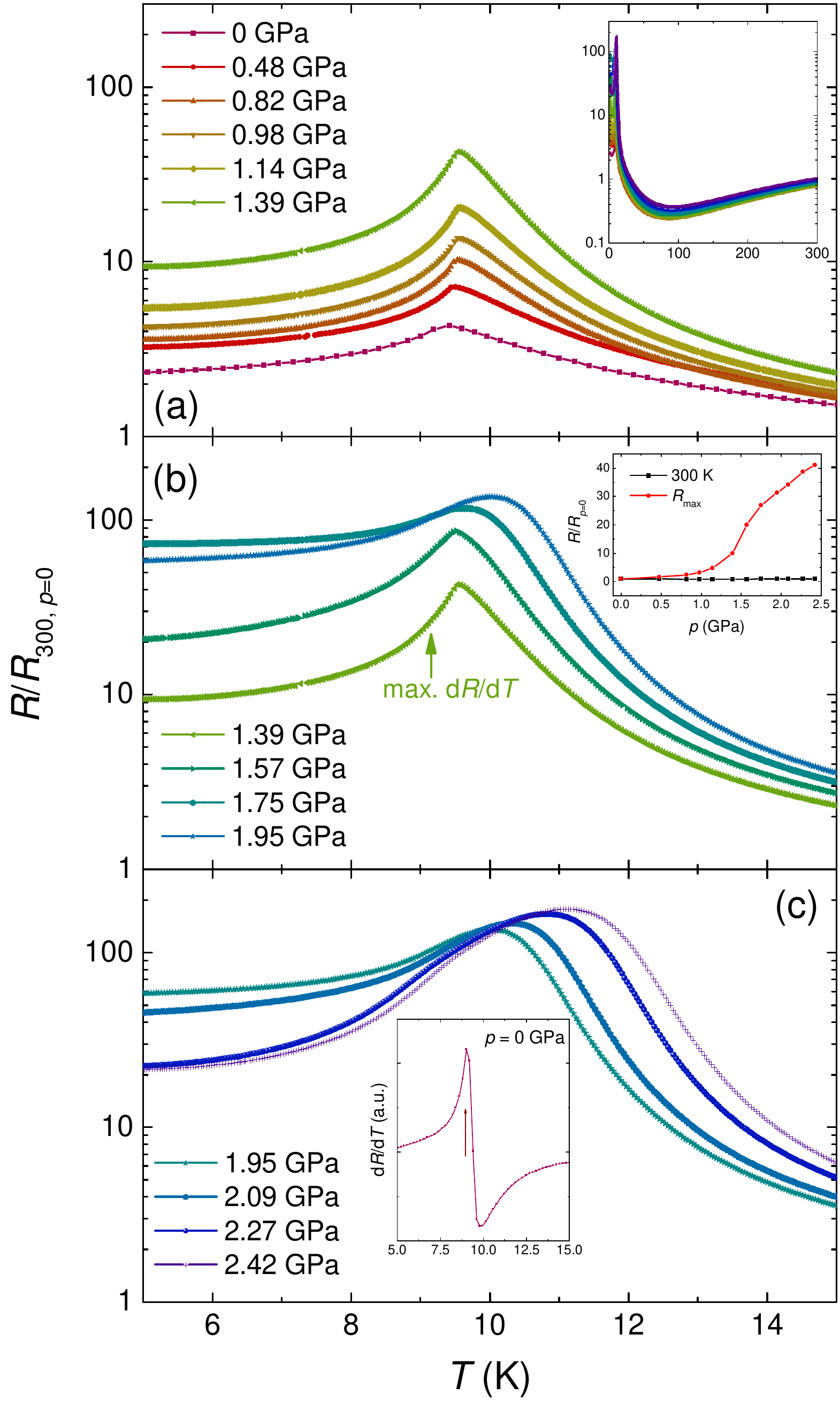} 
\caption{Selected data sets of resistance, normalized to the zero-pressure room-temperature value, $R/R_{300, p=0}$, vs. $T$ (5\,K\,$\le\,T\,\le\,$15\,K) of EuCd$_2$As$_2$ under pressure, $p$, for $0\,$GPa$\,\leq\,p\,\leq\,1.39$\,GPa (a), for $1.39\,$GPa$\,\leq\,p\,\leq\,1.95$\,GPa (b) and $1.95\,$GPa$\,\leq\,p\,\leq\,2.42$\,GPa (c). Current was applied within the $ab$ plane. The inset in (a) shows all data over the full temperature range up to 300\,K. The arrows in (b) indicate the criterion which is used to determine the transition temperature (see text). The inset in (b) depicts the pressure dependence of $R$, normalized to its zero-pressure value $R_{p=0}$, at $T\,=\,300\,$K as well as of $R_\textrm{{max}}.$ The inset in (c) shows an example data set of d$R$/d$T$ at zero pressure, which is used to determine the transition temperatures (see arrow and text).}
\label{fig:resistance}
\end{figure}
\end{center}

To complement our thermodynamic analysis of the phase diagram, we also performed measurements of the resistance, $R$, as a function of temperature at different pressures. The data of $R$, normalized to the room-temperature value at zero pressure $R_{300,p=0}$, is displayed in Fig.\,\ref{fig:resistance} and split into three pressure ranges, similar to the specific heat data. Overall, the $T$-dependent behavior of $R/R_{300,p=0}$ is consistent with the literature at ambient pressure\cite{Wang16,Rahn18,Ma19}. Initially, upon cooling, $R$ decreases, consistent with a semimetallic behavior\cite{Wang16} (see inset of Fig.\,\ref{fig:resistance}\,(a)). Below $\approx\,80\,$K, $R$ starts to increase rapidly \cite{Wang16}. Upon entering the magnetically-ordered state at low temperatures, $R$ decreases due to loss of spin-disorder scattering, resulting in a peak of $R$ at the transition temperature. Over the entire pressure range up to 2.42\,GPa, the peak value of $R$ increases with pressure with the strongest increase observed in the intermediate pressure range 1.4\,GPa\,$\lesssim\,p\,\lesssim\,1.95\,$GPa (see inset of Fig.\,\ref{fig:resistance}\,(b)), while the room-temperature value of $R$ changes by less than 25\,\%. Similar to the specific heat data, we find that the peak in $R$ remains sharp for $p\,\lesssim\,$1.4\,GPa (Fig.\,\ref{fig:resistance}\,(a)), starts to broaden for intermediate $p$ (Fig.\,\ref{fig:resistance}\,b) and remains broad while clearly shifting to higher temperatures beyond 1.95\,GPa (Fig.\,\ref{fig:resistance}\,c).

In order to construct the temperature-pressure phase diagram of EuCd$_2$As$_2$ from the data above, we used the following criteria. For the specific heat data, we determined the position of the minimum in d($C/T$)/d$T$, as exemplarily shown in the inset of Fig.\,\ref{fig:specificheat}\,(b). This criterion is close to the one obtained in isentropic constructions (see arrow in the main panel of Fig.\,\ref{fig:specificheat}\,(b)). For the resistance measurements, we refer to the Fisher-Langer relation \cite{Fisher68} for magnetic transitions in metals and chose the maximum of d$R$/d$T$ (see arrow in Fig.\,\ref{fig:resistance}\,(b) and inset in  Fig.\,\ref{fig:resistance}\,(c)).

The resulting phase diagram is presented in Fig.\,\ref{fig:phasediagram}. Pressure-dependent transition temperature data from both $C$($T$) and $R$($T$) runs agree with each other well and show an initial, slow increase in $T_\textrm{N}$ ($p\,\lesssim\,$1.30\,GPa) with pressure followed by a gradual decrease of $T_\textrm{N}$ with further increase in pressure (1.30\,GPa$\,\lesssim\,p\,\lesssim\,$2.0\,GPa).  For $p\,\gtrsim\,$2.0\,GPa we observe a comparatively sharp increase in the magnetic ordering temperature.  This sharp change in the pressure dependence of the ordering temperature strongly suggests that a different magnetic phase has been stabilized above the critical pressure $p^\star\,\simeq\,2\,$GPa.  Indeed, our $\mu$SR data, shown and discussed below, provide the basis for our determination that the high pressure phase has a ferromagnetic (FM) component to it.

\begin{center}
\begin{figure}
\includegraphics[width=0.9\columnwidth]{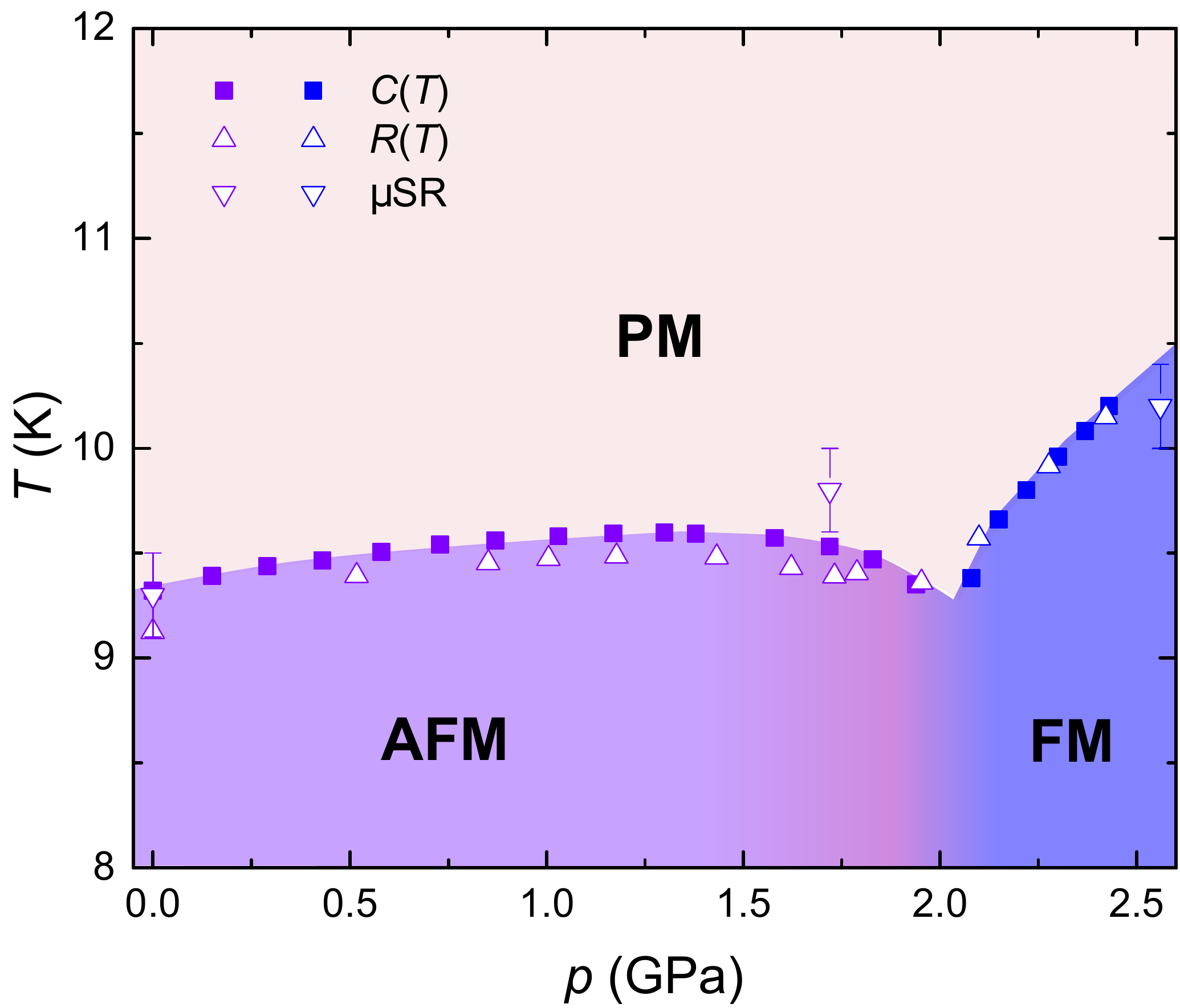} 
\caption{Temperature-pressure ($T$-$p$) phase diagram of EuCd$_2$As$_2$, determined from specific heat, resistance and $\mu$SR measurements. Light purple symbols correspond to the transition from the paramagnetic (PM) to the antiferromagnetic (AFM) state. Blue symbols indicate the position of the PM-ferromagnetic (FM) transition. The color shading in the intermediate pressure range indicates that EuCd$_2$As$_2$ shows strong tendencies towards FM order (see text for more details).}
\label{fig:phasediagram}
\end{figure}
\end{center}

\subsection{Microscopic study of magnetism under pressure}
\label{sec:microscopic-magnetism}

To probe the magnetic properties under pressure, we used $\mu$SR measurements. In these experiments, approximately half of the muons stop in the sample and half in the pressure cell wall. The latter muons are sensitive to the magnetic fields that are generated by the sample inside the pressure cell. Thus, the signal from the muons stopping in the pressure cell wall will be measurably distorted \cite{Khasanov16,Taufour16,Gati21} for ferromagnetic samples. A detailed description of the analysis procedure of the $\mu$SR data under hydrostatic pressure can be found in Appendix Sec.\,\ref{sec:muSRpressureanalysis}.


\begin{center}
\begin{figure}
\includegraphics[width=0.9\columnwidth]{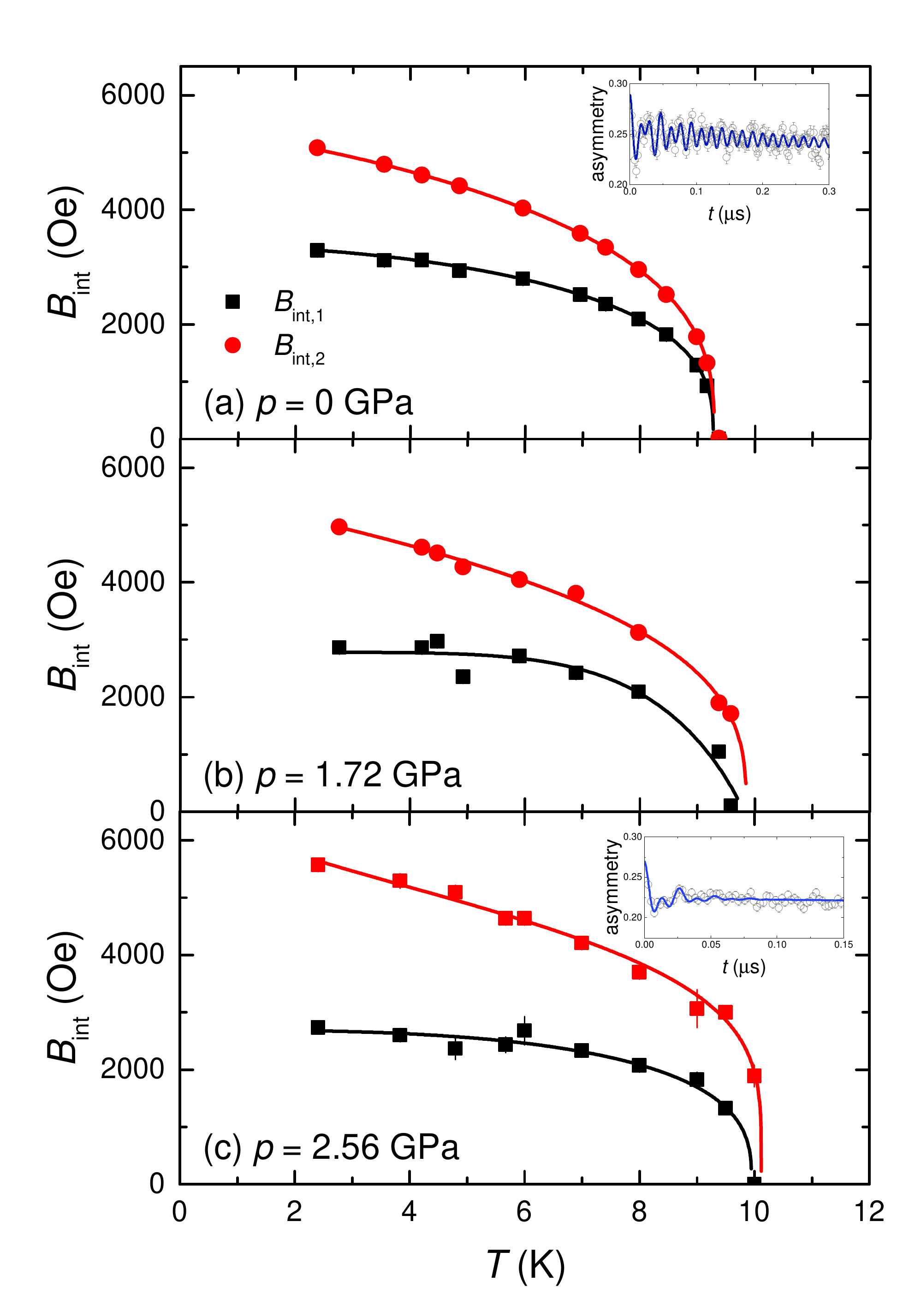} 
\caption{Evolution of the internal field of the two stopping sites in EuCd$_2$As$_2$, $B_{\textrm{int,1}}$ and $B_{\textrm{int,2}}$, with temperature, $T$, for pressures, $p$, of (a) 0\,GPa, (b) 1.72 GPa, and (c) 2.56 GPa. Note that the 0\,GPa data were taken inside the hydrostatic pressure cell. Lines represent fits to the empirical formula, $B_\textrm{int,i}\,=\,B_\textrm{int,i}^0 \left[1-\left(\frac{T}{T^\star}\right)^\alpha \right]^\beta$ (see text). The insets in (a) and (b) show zero-field spectra (open grey symbols) taken at $T\,=\,3.55\,$K and 2.41\,K, respectively, together with the fit for two stopping sites (blue lines).}
\label{fig:internalfield}
\end{figure}
\end{center}

First, we present in Fig.\,\ref{fig:internalfield} the evolution of the internal field of EuCd$_2$As$_2$ with temperature and pressure, which can be inferred from the zero-field $\mu$SR spectra (see insets in (a) and (c) for the asymmetry data at 0\,GPa and 2.56\,GPa). In total, we took high-statistics zero field data sets for 0\,GPa (Fig.\,\ref{fig:internalfield}\,(a)), 1.72\,GPa (Fig.\,\ref{fig:internalfield}\,(b)) and 2.56\,GPa (Fig.\,\ref{fig:internalfield}\,(c)). For all pressures, clear oscillations were observed below the respective pressure-dependent ordering temperature, signalling the presence of a finite internal field below the respective transition temperature. In each case, our data is best modeled by using two muon stopping sites with internal fields $B_\textrm{int,1}$ and $B_\textrm{int,2}$. For the 0\,GPa data, the best fit was obtained by using a ratio of 60:40 for the two stopping sites, whereas for higher pressures we fixed the ratio to 50:50. The transition temperatures from $\mu$SR, which are included in the phase diagram in Fig.\,\ref{fig:phasediagram}, were determined by fitting the $B_\textrm{int,i}$ data with the empirical formula $B_\textrm{int,i}^0 \left[1-\left(\frac{T}{T^\star}\right)^\alpha \right]^\beta$, with $B_\textrm{int,i}^0$, $T^\star$, $\alpha$ and $\beta$ being free parameters ($T^\star$ corresponds to the respective transition temperature). Interestingly, the ratio of internal fields $B_\textrm{int,2}/B_\textrm{int,1}$ at $T\,\approx\,2\,$K increases significantly with pressure from 1.5 at 0\,GPa to 1.7 at 1.72\,GPa to 2 at 2.56\,GPa. Even though we lack precise information on the muon stopping sites in EuCd$_2$As$_2$, the change of the ratio of internal fields suggests that there is some change of the magnetic structure, for example by rotation of the moments at a muon stopping site, with pressure \cite{Khasanov17}.

\begin{center}
\begin{figure}
\includegraphics[width=0.9\columnwidth]{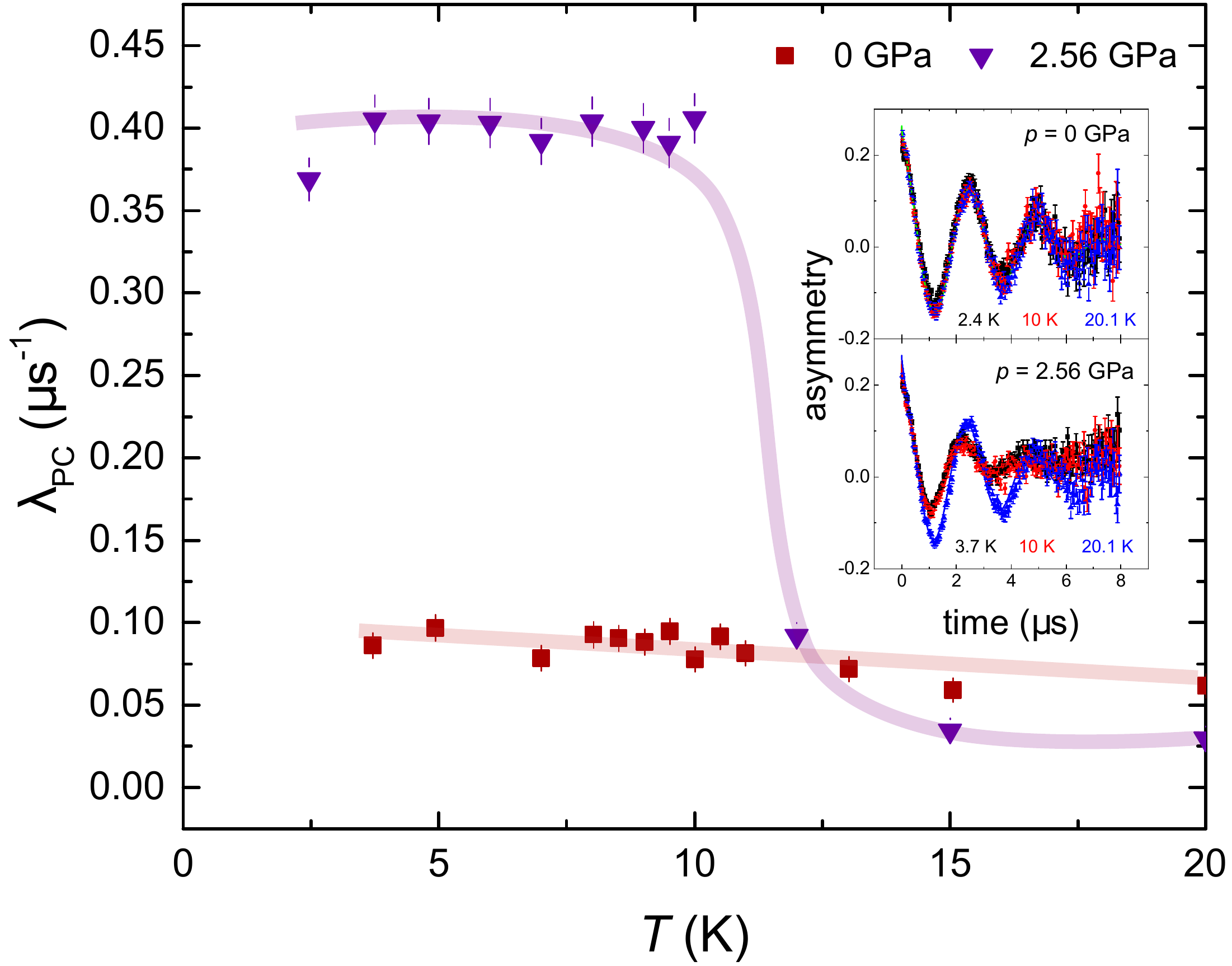} 
\caption{Temperature dependence of the relaxation rate of the pressure cell, $\lambda_{\textrm{PC}}$, for pressures of 0\,GPa and 2.56\,GPa. Lines are guide to the eyes. The insets the weak-transverse field spectra at 0\,GPa and at 2.56 GPa for temperatures below and above the phase transition.}
\label{fig:pressurecellresponse}
\end{figure}
\end{center}

Next, we discuss our results of the type of magnetic order under pressure. To this end, we show in Fig.\,\ref{fig:pressurecellresponse} the pressure cell relaxation rate, $\lambda_\textrm{PC}$, at two different pressures, which were obtained from measuring the muon asymmetry in a weak transverse field of 30\,Oe. As already evident from the raw data shown in the insets, there is no additional depolarization below the ordering temperature $T_\textrm{N}\,\approx\,9.5$\,K at ambient pressure ($p\,<\,p^\star$), whereas there is a clear additional depolarization below $T_\textrm{C}\,\approx\,10.4\,$K at 2.56\,GPa ($p\,>\,p^\star$). As depicted in the main panel, the former data correspond to a weak temperature dependence of $\lambda_\textrm{PC}$, the size of which equals the known relaxation rate of the pressure cell \cite{Khasanov16} (see Appendix Sec.\,\ref{sec:muSRpressureanalysis}). This behavior is expected for antiferromagnetic order (see also our discussion of ambient-pressure $\mu$SR data on oriented single crystals in Appendix Sec.\,\ref{sec:muSRambient}). In contrast, for 2.56\,GPa, $\lambda_\textrm{PC}$ suddenly increases below $T_\textrm{C}$ and levels off for lower temperatures, which can only be attributed to the presence of an additional field created by the sample. Importantly, this central result of our work represents compelling evidence  for the realization of a FM state at high pressures. This conclusion is further supported by an increase of ferromagnetic fluctuations above the ordering temperature with pressure (see Appendix Sec.\,\ref{sec:muSRrelaxation}). Based on the presented $\mu$SR data on an aggregate of randomly oriented single crystals, we cannot make any statement on whether the FM state is fully polarized along a crystallographic direction or whether there is only a ferromagnetic component to the order. Below, based on anisotropic magnetoresistance data, we argue that the high-pressure FM state is likely almost in-plane polarized, and thus we refer here to the notion of FM order.


\subsection{Anisotropic magnetoresistance data}
\label{sec:anisotropic-magnetoresistance}

Now that we have established a change of the magnetic ground state from AFM to FM by hydrostatic pressure in EuCd$_2$As$_2$, we want to discuss transport measurements under pressure in finite magnetic field in the ordered states. The aim of these experiments is to identify the moment direction in the high-pressure state.

\begin{center}
\begin{figure}
\includegraphics[width=1\columnwidth]{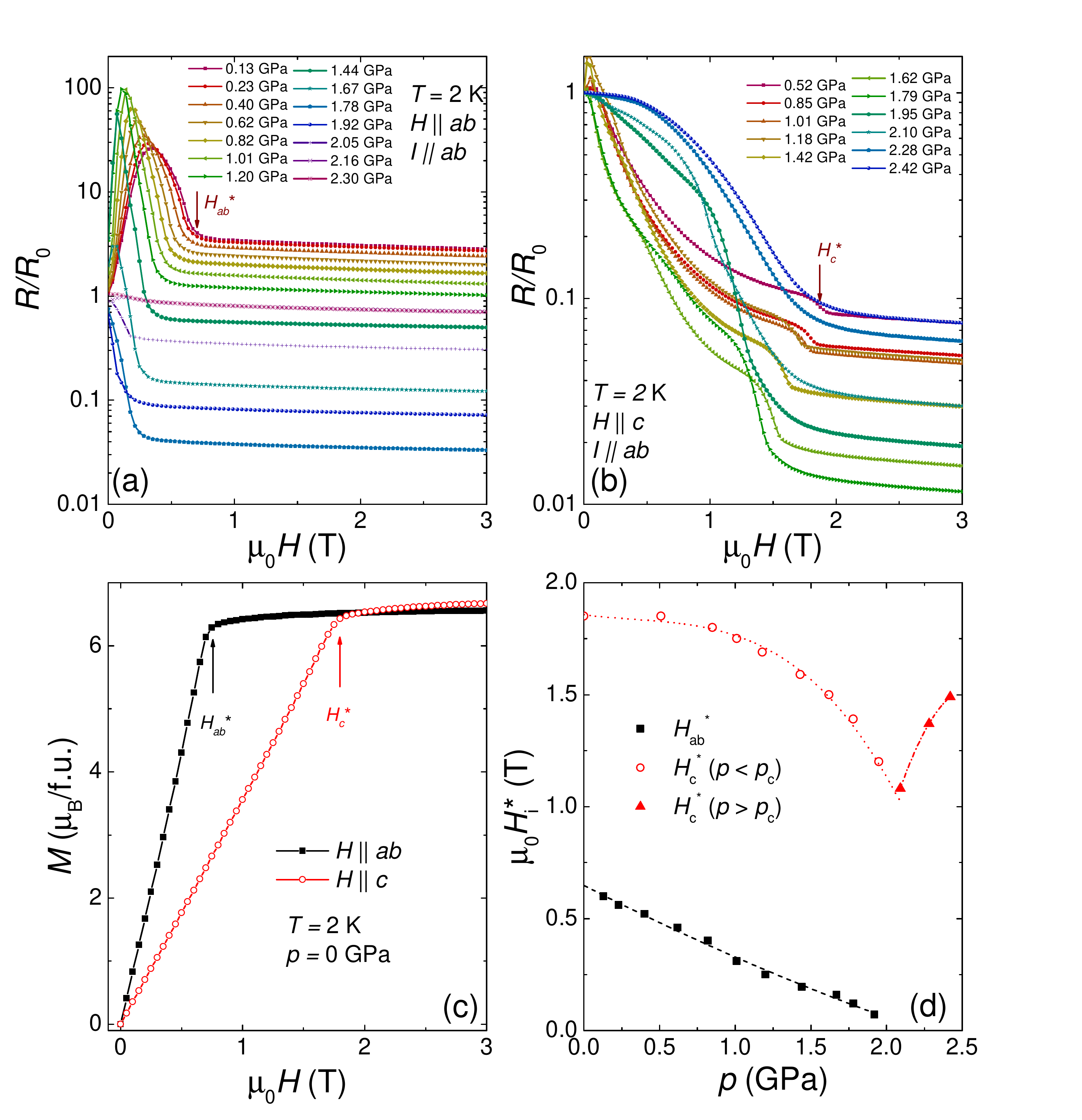} 
\caption{Anisotropic magnetoresistance of EuCd$_2$As$_2$ at $T\,=\,2\,$K at different pressures. (a) In-plane $R/R_0$, with $R_0$ being the resistance in zero external field, as a function of in-plane field $\mu_0H$ at different pressures $p$. Arrow indicates the criterion to determine $H_\textrm{ab}^\star$; (b) In-plane $R/R_0$ as a function of out-of-plane field $\mu_0H$ at different pressures $p$. Arrow indicates the criterion to determine $H_\textrm{c}^\star$; (c) Ambient-pressure anisotropic magnetization as a function of in-plane  as well as out-of-plane magnetic field \cite{Jo20}. Arrows indicate the saturation fields $H_\textrm{ab}^\star$ and $H_\textrm{c}^\star$; (d) Evolution of $H_\textrm{ab}^\star$ and $H_\textrm{c}^\star$ with pressure. For $p\,<\,p_c$, the ground state is AFM with moments lying in the $ab$-plane. Thus, $H_\textrm{ab}^\star$ and $H_\textrm{c}^\star$  are defined as saturation fields into fully polarized states with moments in the $ab$ and $c$ direction, respectively. For $p\,>\,p_c$, the ground state is FM with moments in the $ab$ plane, thus only $H_\textrm{c}^\star$ can be defined.}
\label{fig:MR}
\end{figure}
\end{center}

In Figs.\,\ref{fig:MR} (a) and (b), we show our data of the pressure-dependent magnetoresistance (MR) data at $T\,=\,2\,$K, taken in a longitudinal and in a transverse configuration, i.e., $\mu_0 H \parallel I \parallel ab$,  (a) and $\mu_0 H \parallel c, I \parallel ab$ (b). At lowest pressures (0.13\,GPa and 0.52\,GPa, respectively), the longitudinal $R/R_0$ goes through a maximum and becomes almost field-independent above $\mu_0 H^\star_\textrm{ab}\,\sim\,$0.7\,T (Fig.\,\ref{fig:MR}\,(a)), whereas the transverse $R/R_0$ shows a step-like change around $\mu_0 H^\star_\textrm{c}\,\sim\,$1.8\,T (c,d). Those field scales coincide very well with the saturation fields for the specific field orientations, determined from magnetization measurements at ambient pressure \cite{Jo20} (see Fig.\,\ref{fig:MR}\,(c)).

As clearly visible in the raw data (a), $H^\star_\textrm{ab}$ decreases rapidly with increasing pressure. As shown in Fig.\,\ref{fig:MR}\,(d), $H^\star_\textrm{ab}$ extrapolates to zero at $\approx\,$2\,GPa, i.e., at $p_c$. For 2.3\,GPa, the $R/R_0$ data is essentially field-independent. These results strongly suggest that the high-pressure FM state has its moments aligned in the plane, i.e., a pure FM state: With pressure, we suppress $H^\star_\textrm{ab}$ to zero, implying full moment alignment along the in-plane direction.

Whereas $H^\star_\textrm{ab}$ is suppressed to zero with increasing pressure, $H^\star_\textrm{c}$ is also suppressed, but remains finite at 2\,GPa. For $p\,\gtrsim\,$2\,GPa, no step-like change can be observed. Instead, we observe a broad crossover feature in the field-dependent $R/R_0$ data, the position of which increases with increasing pressure (see Fig.\,\ref{fig:MR}\,(d)). The absence of a metamagnetic transition for out-of-plane fields is consistent with the notion of FM order with moments oriented in the $ab$-plane. The decrease in $H^\star_\textrm{c}$ for $p\,<\,p_c$ is nonetheless remarkable, as it suggests that the energy difference to the FM state with moments along the $c$ axis is also reduced with pressure. We will further discuss the proximity to this FM$_\textrm{c}$ state based on DFT calculations below. We note that our Hall data under pressure, presented in Appendix Sec.\,\ref{sec:Hall}, indicate the possibility that the change of magnetic structure across $p_c$ is associated with a change of charge carrier density.

\subsection{Results of DFT calculations under pressure}
\label{sec:DFTcalculations}

\begin{center}
\begin{figure}
\includegraphics[width=0.9\columnwidth]{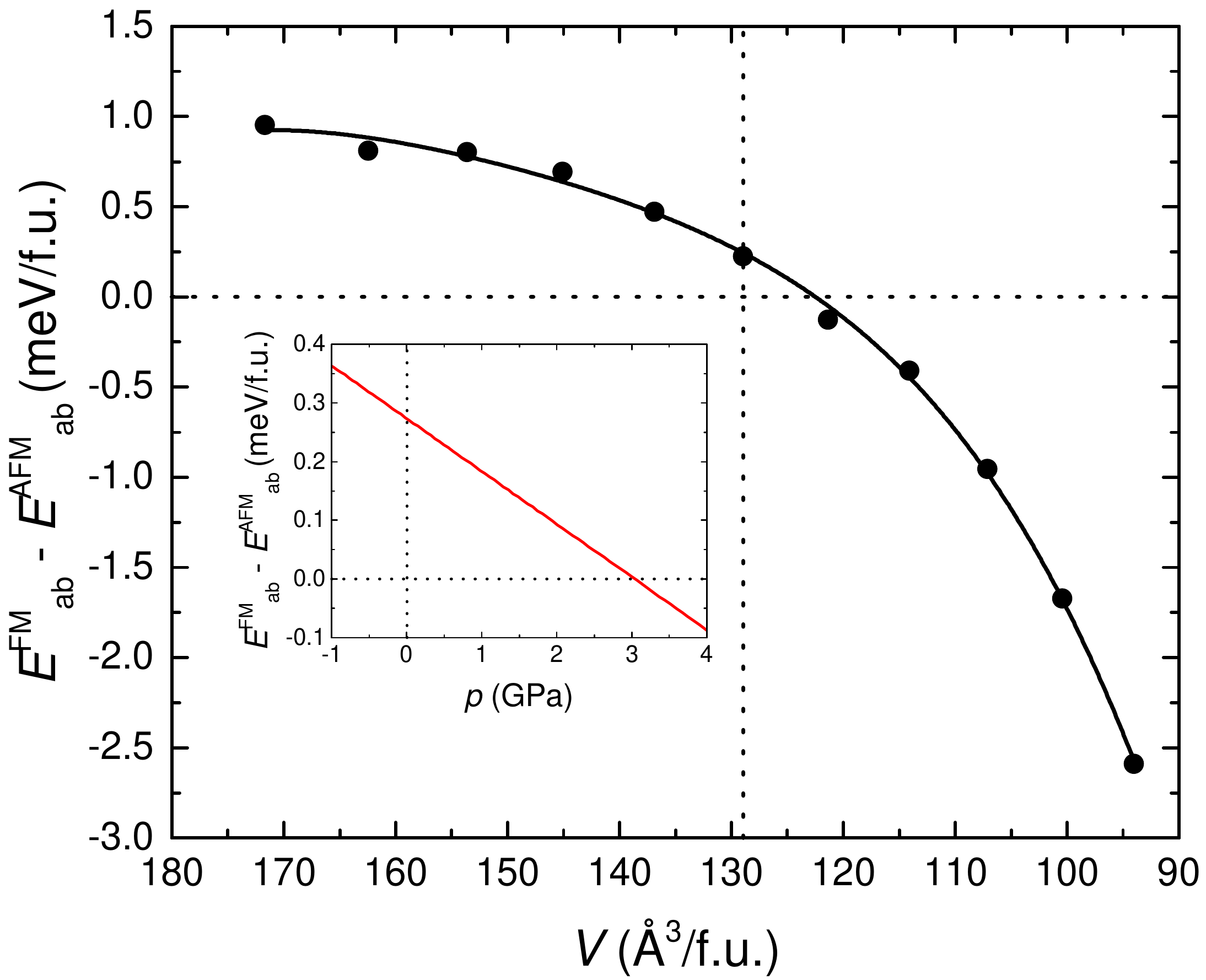} 
\caption{{Equation of state (EOS) energy difference ($\Delta E_{FM,ab-AFM,ab}$) vs. volume ($V$) between the fully-relaxed FM and A-type AFM, where in both cases the moment point in-plane to the nearest-neighbor Eu. The initial AFM$_{ab}$ equilibrium volume of 128.96 \AA$^3$/f.u. as well as the phase transition at $\Delta E_{FM,ab-AFM,ab}$=0.0 are indicated by the vertical and horizontal dashed line, respectively. The inset shows an enlarged view around the AFM$_{ab}$-FM$_{ab}$ transition region with the \textit{x}-axis converted to pressure ($p$) using the calculated bulk modulus (see text). The dashed line for $\Delta E_{FM,ab-AFM,ab}$=0.0 gives the critical pressure $p_c\,\approx\,$3.0 GPa.}}
\label{fig:EOS}
\end{figure}
\end{center}

In this section we study the pressure-dependent evolution of magnetic order in EuCd$_2$As$_2$ and the possible transition from in-plane AFM (denoted by AFM$_{ab}$ in the following) to in-plane FM order (denoted by FM$_{ab}$) for low pressures with the help of DFT calculations. To this end, we constructed the EOS for both AFM$_{ab}$ and FM$_{ab}$ spin configurations around the equilibrium volume and fitted it to the Birch-Murnaghan equation\cite{Birch47,Murnaghan44}. The optimized equilibrium lattice constants using the PBE exchange-correlation functional and $U$\,=\,4.4\,eV without spin-orbit coupling (SOC) for A-type antiferromagnetic (i.e., AFM$_{ab}$) spin configuration are $a\,$=\,4.49\,\AA~ and $c$\,=\,7.37\,\AA. These values agree with the experimental data of 4.4336 and 7.2925 \AA\, at $T\,=\,11$\,K respectively, just above $T_N$ (see Appendix Sec.\,\ref{sec:xraydiffraction}) within about 1\%, which can be considered as a very good agreement for DFT-PBE calculations \cite{Harl10}. For EOS calculations, at each volume, the cell shape and atomic positions are fully relaxed for both AFM$_{ab}$ and FM$_{ab}$ with SOC, with the magnetic moment pointing in the in-plane direction to the nearest-neighbor (NN) Eu or equivalently along $a$- or $b$-axis. The fitted equilibrium volumes from EOS are 129.78 and 129.76 \AA$^3$/f.u. for AFM$_{ab}$ and FM$_{ab}$ with SOC, respectively, having an energy difference of 0.3 meV/f.u. in favor of the AFM$_{ab}$ state. This agrees well with the 0.22 meV/f.u. preference to AFM$_{ab}$ from the direct DFT calculation at the initial AFM equilibrium volume of 128.96 \AA$^3$/f.u. With such small differences in equilibrium volume and energy, the two fully-relaxed EOS of the competing magnetic phases are aligned very closely together to have almost the same bulk modulus of 46.7\,GPa. To obtain the critical pressure $p_c$ for the AFM-FM magnetic phase transition, the difference between the two EOS, $\Delta E_{FM,ab-AFM,ab}$ vs. volume is plotted in Fig.\ref{fig:EOS}. At the initial AFM$_{ab}$ equilibrium volume, AFM$_{ab}$ is preferred by 0.22\,meV/f.u. This preference is increased at larger volume, but decreased for smaller volume, i.e., with increasing hydrostatic pressure. This leads to a AFM$_{ab}$-FM$_{ab}$ magnetic phase transition under hydrostatic pressure, as observed in experiment. The transition region is shown on enlarged scales in the inset of Fig.\ref{fig:EOS} with the \textit{x}-axis converted to pressure using the calculated bulk modulus. The dashed line of $\Delta E_{FM,ab-AFM,ab}$\,=\,0.0 gives the critical pressure $p_c\,\approx\,$3.0 GPa, in good agreement with the experimental data of $\approx\,$2.0\,GPa. To firmly establish the pressure-induced AFM$_{ab}$-FM$_{ab}$ in EuCd$_2$As$_2$ within DFT, we have also tested different exchange-correlation functionals, PAW potentials and computational settings, including using the same relaxed crystal structures for AFM$_{ab}$ and FM$_{ab}$ at each volume from either the non-magnetic or AFM$_{ab}$ configuration. All calculations confirm a  AFM$_{ab}$-FM$_{ab}$, with $p_c$ values in the range from 2.7 to 4.2\,GPa (see App. \ref{app:lattice-parameters-theory} for a detailed discussion on different initial conditions).

Next, we calculated the crystalline magneto-anisotropy energy (MAE), defined by the required energy to switch the crystal magnetization from the low pressure basal plane magnetic order and direction to, in this case explicitly, FM$_c$ order, at increasing pressure. Such a calculation provides information about the proximity of EuCd$_2$As$_2$ to the FM$_c$ state, which, according to recent work, is able to host a single pair of Weyl nodes \cite{Hua18,Wang19}. This calculation was done in various steps. First, we relaxed the unit cell of EuCd$_2$As$_2$ at different volumes by treating the Eu$^{2+}$ 4f states  completely  as  core  states  and by performing non-spin polarized calculations with  the PBE  exchange  correlation  functional. The former warrants that no spurious Eu f states appear near the Fermi level and the latter ensures a fully  non-magnetic  setting  since no magnetic moments/spins are invoked in the calculation. Such a relaxation guarantees a uniform change in the lattice without being influenced by the nature of the assumed magnetism for the Eu$^{2+}$ 4f states. The resulting optimized non-magnetic equilibrium lattice parameters are given by $a\,=\,4.48\,$\,\AA and $c\,=\,7.27\,$\,\AA, which are closer to the
experimental data (see Appendix \ref{sec:xraydiffraction}) than the magnetically relaxed structures, due to the absence of long-range magnetic order at the measured $T\,=\,11\,$K. In a second step, we took the optimized non-magnetic crystal structures and calculated  total energies for different Eu magnetic configurations, by including the Eu$^{2+}$ 4f states as valence electrons with $U\,=\,4.4\,$eV and SOC and with the Eu moments aligned either antiferromagnetically or
ferromagnetically and moments pointing in the plane for AFM$_{ab}$ and FM$_{ab}$, respectively, as well as aligned ferromagnetically along the $c$-axis, denoted as FM$_c$. The DFT results were fitted to the Birch-Murnagham EOS, yielding the MAE by subtracting the corresponding fits. After fitting, the equilibrium volume increases from $126.4$\,\AA$^{3}/$f.u.~ for the non-magnetic configuration to  $130.21$\,\AA$^{3}/$f.u.~for both FM orders and to $130.21$\,\AA$^{3}/$f.u.~for AFM$_{ab}$. Note that throughout all performed calculations we keep the valence of Eu fixed to 2+, so that we cannot exclude the possibility of a valence transition (Eu$^{2+}$ to Eu$^{3+}$) at the simulated higher pressures. This goes beyond the scope of the present work. 

\begin{center}
\begin{figure}
\includegraphics[width=1.0\columnwidth]{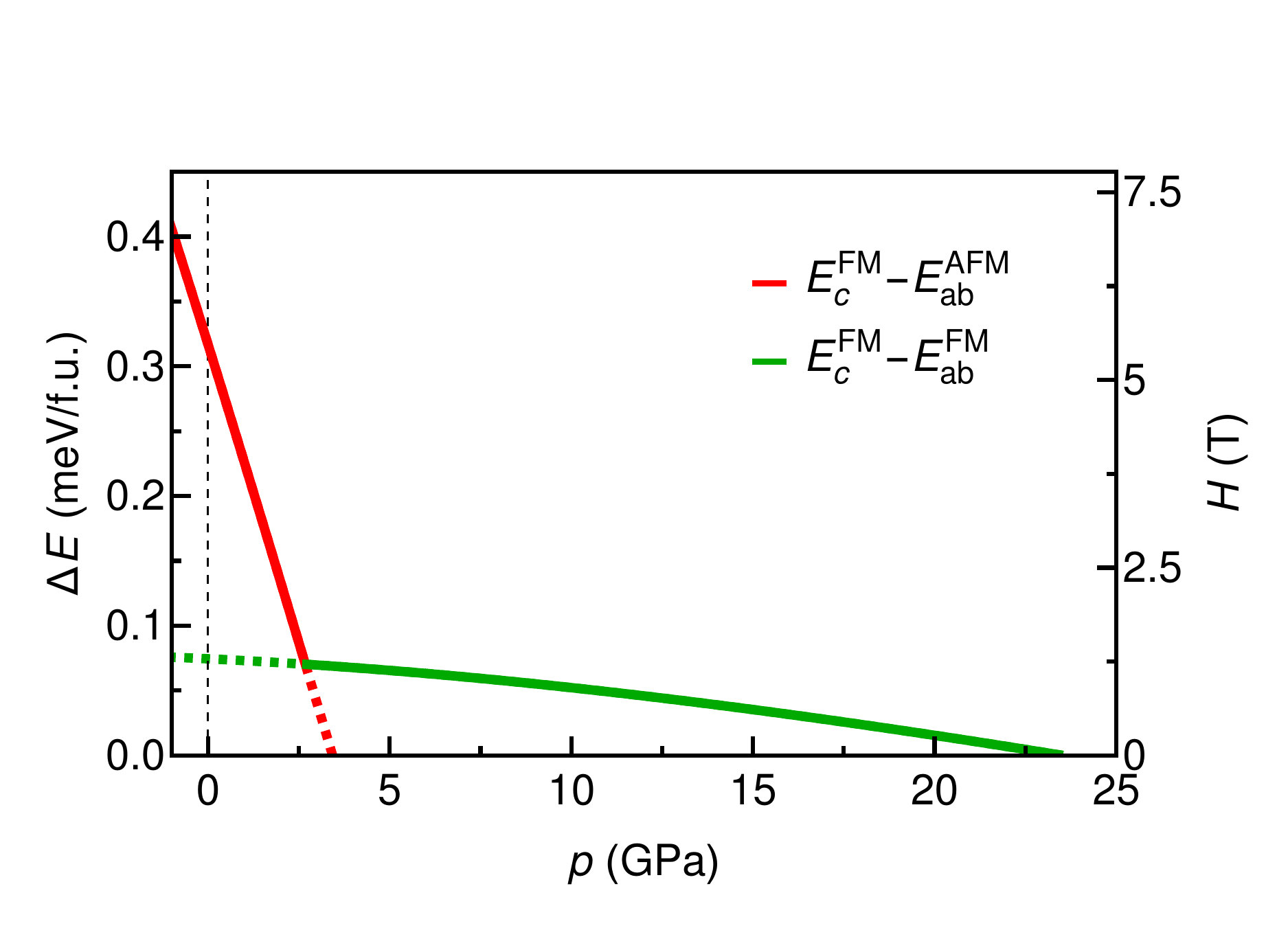} 
\caption{{Energy difference of the FM$_{c}$ order to the ground states (left axis) and corresponding magnetic field (right axis) vs. pressure. The solid lines show the energy difference between the FM$_c$ state  and the respective ground states, i.e., AFM$_{ab}$ for $p\,\lesssim\,$3\,GPa (red) and FM$_{ab}$ for $p\,\gtrsim\,$3\,GPa (green). Dashed lines represent the energy differences between non-ground state magnetic configurations. The ground state transition to FM$_c$ takes place at approximately $p'_c \approx 23$ GPa.}}
\label{fig:NM-rlx}
\end{figure}
\end{center}

In Fig.~\ref{fig:NM-rlx} we show the results of these MAE calculations as a function of hydrostatic pressure. The MAE is directly proportional to the $H_c^\star$ field that is measured in experiment. At ambient pressure, the theoretical calculations overestimate the experimental $H_c^\star$ value, which is expected in the framework of DFT and Birch-Munagham EOS, which we use. Nevertheless, the tendencies as a function of pressure are robust. Here, we find that the MAE or $H_c^\star$ decrease rapidly with increasing pressure for $p\,<\,p_c$, consistent with our experimental observations, shown in Fig.\,\ref{fig:MR}\,(d). Above $p_c$, the MAE still continues to decrease with increasing pressure, however at a slower rate. This trend is opposite to what is observed for our experimental data for pressures close to $p_c$. However, we note that the details of our theoretical prediction close to $p_c$ depend on the details of the relaxation (The result of a different relaxation is shown in Fig.\,\ref{fig:AF-rlx}). Thus, experimental data to higher pressures is needed for a meaningful comparison between theory and experiment in the FM$_{ab}$ ground state region.

The decrease of MAE with pressures beyond $\approx\,10$\,GPa does not depend on details of the relaxation. Thus, there will eventually be a transition to the FM$_c$ state at sufficiently high pressures, as long as Eu does not change its valency. We predict that this transition to an FM$_c$ state occurs at $p'_c\approx23$ GPa. Whereas this pressure is larger than the experimentally-used pressures in this work, it is nevertheless experimentally feasible to realize those pressures. Thus, based on our theoretical calculations, it might be possible to stabilize the desired FM$_c$ state in EuCd$_2$As$_2$ in experiment as a result of pressure effects and without application of an external magnetic field, and, to study key transport quantities with respect to topology. At the same time, it will be very interesting to perform further studies to understand the microscopic origin of the high tunability of the magnetic properties of EuCd$_2$As$_2$.


\section{Conclusions}
\label{sec:conclusions}

In summary, we presented an experimental study of the temperature-pressure phase diagram up to $\sim\,2.5$\,GPa of the magnetic Weyl semimetal EuCd$_2$As$_2$ by means of specific heat, transport and $\mu$SR measurements and performed DFT calculations up to pressures of $\approx$\,25\,GPa. Our results clearly show that this compound undergoes a transition from an antiferromagnetic state at low pressures to a ferromagnetic state above 2\,GPa, in which moments are dominantly oriented within the $ab$ plane. In addition, we find that pressure also drives EuCd$_2$As$_2$ closer to the sought-for FM state, in which moments are oriented along the $c$ axis. We predict that a hydrostatic pressure of $\approx$\,23\,GPa will stabilize this FM$_c$ state, if no valence transition of Eu$^{2+}$ intervenes. Overall, our study identifies a clear and experimentally-accessible tuning route to change the magnetic ground state in EuCd$_2$As$_2$ in a single sample
and investigate the interplay of magnetism and topological non-trivial phases. Our results motivate further experimental and theoretical studies on EuCd$_2$As$_2$ under pressure, which is a very powerful parameter to tune the properties of this magnetic Weyl semimetal. 

\section{Acknowledgments}
We thank Na Hyun Jo and Young-Joon Song for useful discussions. SLB and PCC thank B. Schweinsteiger for inspiring research group members. Work at the Ames Laboratory was supported by the U.S. Department of Energy, Office of Science, Basic Energy Sciences, Materials Sciences and Engineering Division. The Ames Laboratory is operated for the U.S. Department of Energy by Iowa State University under Contract No. DEAC02-07CH11358. E.G. and L.X. were funded, in part, by the Gordon and Betty Moore Foundation's EPiQS Initiative through Grant No. GBMF4411. B.K. and L.-L.W. were funded by the Center for Advancement of Topological Semimetals, an Energy Frontier Research Center funded by the U.S. Department of Energy Office of Science, Office of Basic Energy Sciences, through the Ames Laboratory under its Contract No. DE-AC02-07CH11358. Research of R.G. is supported by the Swiss National Science Foundation (SNF-Grant No. 200021-175935). AV and RV acknowledge support by the Deutsche Forschungsgemeinschaft (DFG, German Research Foundation) for funding through TRR 288 - 422213477 (project B05).

\section{Appendix}
\label{sec:Appendix}

\subsection{Additional discussion of $\mu$SR data under pressure}

\subsubsection{Analysis of $\mu$SR data inside a pressure cell}
\label{sec:muSRpressureanalysis}

\textit{Zero field data analysis -} The zero-field $\mu$SR data was analyzed by taking two independent contributions to the total asymmetry, $A(t)$, into account

\begin{equation}
A^\textrm{ZF}(t) = A^\textrm{ZF}_\textrm{S}(t) + A^\textrm{ZF}_\textrm{PC}(t),
\end{equation}

with $A^\textrm{ZF}_\textrm{S}(t)$ ($A^\textrm{ZF}_\textrm{PC}(t)$) being the contribution from the sample (the pressure cell). The sample contribution for EuCd$_2$As$_2$ was found to be best described by considering two stopping sites with different internal field, resulting in

\begin{equation}
\begin{split}
&A^\textrm{ZF}_\textrm{S}(t) = \\\  &A^\textrm{ZF}_\textrm{S,0} [ \sum_{i=1}^{2}m_i \left( \frac{2}{3} \cos{(\gamma_\textrm{m} B_\textrm{int,i}) \exp{(-\lambda_\textrm{T,i})} + \frac{1}{3} \exp{(-\lambda_\textrm{L,i})}} \right) \\\ &+(1-m_1-m_2) \exp{(-\Lambda t)} ]
\end{split}
\end{equation}

with $m_i$ the magnetic fraction of the $i$-th component, $\gamma_\textrm{m}$ the gyromagnetic ratio of the muon, $B_\textrm{int,i}$ the internal field, $\lambda_\textrm{T,i}$ the transverse relaxation rate, $\lambda_\textrm{L,i}$ the longitudinal relaxation rate, and $\Lambda$ the relaxation rate of the paramagnetic portion of the sample. The 2/3 and 1/3 components arise from averaging over a large aggregate of arbitrarily-oriented single crystals. 

The background contribution from the pressure cell can be determined in an independent set of experiment and can be described by two depolarization channels (one originating from nuclear moments and one from electronic moments) following a damped Kubo-Toyabe form

\begin{equation}
\begin{split}
&A^\textrm{ZF}_\textrm{PC}(t) = \\\ & A^\textrm{ZF}_\textrm{PC,0} \left( \frac{1}{3} + \frac{2}{3}(1-\sigma_\textrm{PC}^2t^2) \right) \exp{(-\sigma_\textrm{PC}^2t^2/2)} \exp{(-\lambda_\textrm{PC} t)},
\end{split}
\end{equation}

 with $\sigma_\textrm{PC}$ ($\lambda_\textrm{PC}$) the relaxation rate associated with the nuclear (electronic) moments. 
 
 \textit{Weak transverse field data analysis -} As mentioned in the main text, there exists an additional depolarization of the muons stopping in the pressure cell when the sample inside the cell exhibits a strong magnetization. We note that the muons which stop in the pressure cell wall mostly stop inside the inner cylinder, i.e., very close to the sample, so that they can be sensitive to stray fields emerging from the sample, as shown in recent calculations of the muon stopping profile for pressure-cell experiments with this particular muon energy \cite{Shermandini17}. In the case of weak-transverse field experiments with finite $B_\textrm{ex}$, the pressure cell contribution thus reads as
 
 \begin{equation}
\begin{split}
&A^\textrm{wTF}_\textrm{PC}(t) = \\\ & A^\textrm{wTF}_\textrm{PC,0}(t) \exp(-\sigma_\textrm{PC}^2 t^2/2) \exp(-\lambda_\textrm{PC}) \cos(\gamma_\textrm{m} B_\textrm{ex} t + \phi).    
\end{split}
\end{equation}  

Here, $\sigma_\textrm{PC}$ refers to the relaxation rate caused by nuclear moments, whereas $\lambda_\textrm{PC}$ is relaxation rate determined by the electronic moments as well as the influence of the field that is created by a sample with macroscopic magnetization. The electronic contribution to $\lambda_\textrm{PC}$ is typically almost temperature-independent\cite{Khasanov16} and $\lambda_\textrm{PC}\,\simeq\,0.05-0.1\,\mu$s  (see also data above the ordering temperature in Fig.\,\ref{fig:pressurecellresponse}).

\subsubsection{Evolution of $mu$SR relaxation above the ordering temperature with pressure}
\label{sec:muSRrelaxation}

\begin{center}
\begin{figure}
\includegraphics[width=0.9\columnwidth]{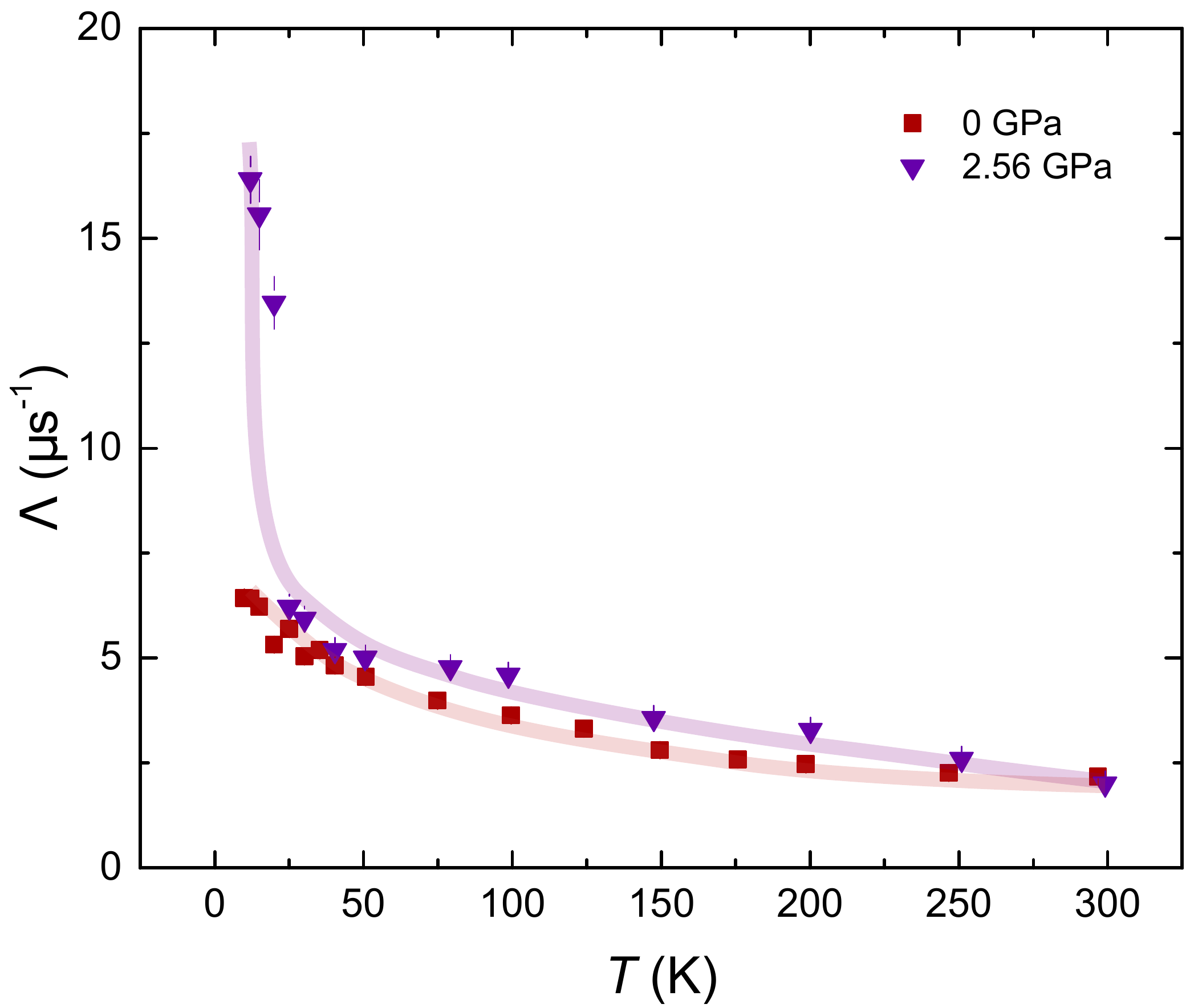} 
\caption{Temperature dependence of the dynamical relaxation rate, $\Lambda$, of EuCd$_2$As$_2$ for pressures of 0\,GPa and 2.56\,GPa for $10\,$K$\,<\,T\,<\,300$\,K and $12\,$K$\,<\,T\,<\,300$\,K, respectively, i.e., in the paramagnetic state at each pressure. Lines are guide to the eyes.}
\label{fig:fluctuations}
\end{figure}
\end{center}

We now discuss the relaxation rate, $\Lambda$, of EuCd$_2$As$_2$ above the ordering temperature, as shown in Fig.\,\ref{fig:fluctuations}. For ambient pressure, we find a moderate increase of $\Lambda$ with lowering temperature, which is fully consistent with the expectations for paramagnetism stemming from the large Eu$^{2+}$ moments. Importantly, we note that we do not find any indications for the step-like change of $\Lambda$ at $\approx\,100$\,K, which Ma \textit{et al.} \cite{Ma19} observed and which was taken as a strong evidence for quasi-long range and quasi-static magnetic order below 100\,K that stabilizes Weyl physics even without spontaneous breaking of time-reversal symmetry (see Appendix \ref{sec:muSRambient} for supporting $\Lambda$ data on a well-oriented single crystal at ambient pressure outside the pressure cell). Instead, the large contribution from the Eu paramagnetism does not allow for any conclusion about the presence or nature of magnetic correlations. However, upon increasing pressure to $p\,=\,2.56$\,GPa, we find that $\Lambda$ increases much faster below $\approx\,$20\,K, reflecting a strong enhancement of FM fluctuations in the proximity of the FM ground state.

\subsubsection{$\mu$SR data at ambient pressure}
\label{sec:muSRambient}

\begin{center}
\begin{figure}
\includegraphics[width=0.9\columnwidth]{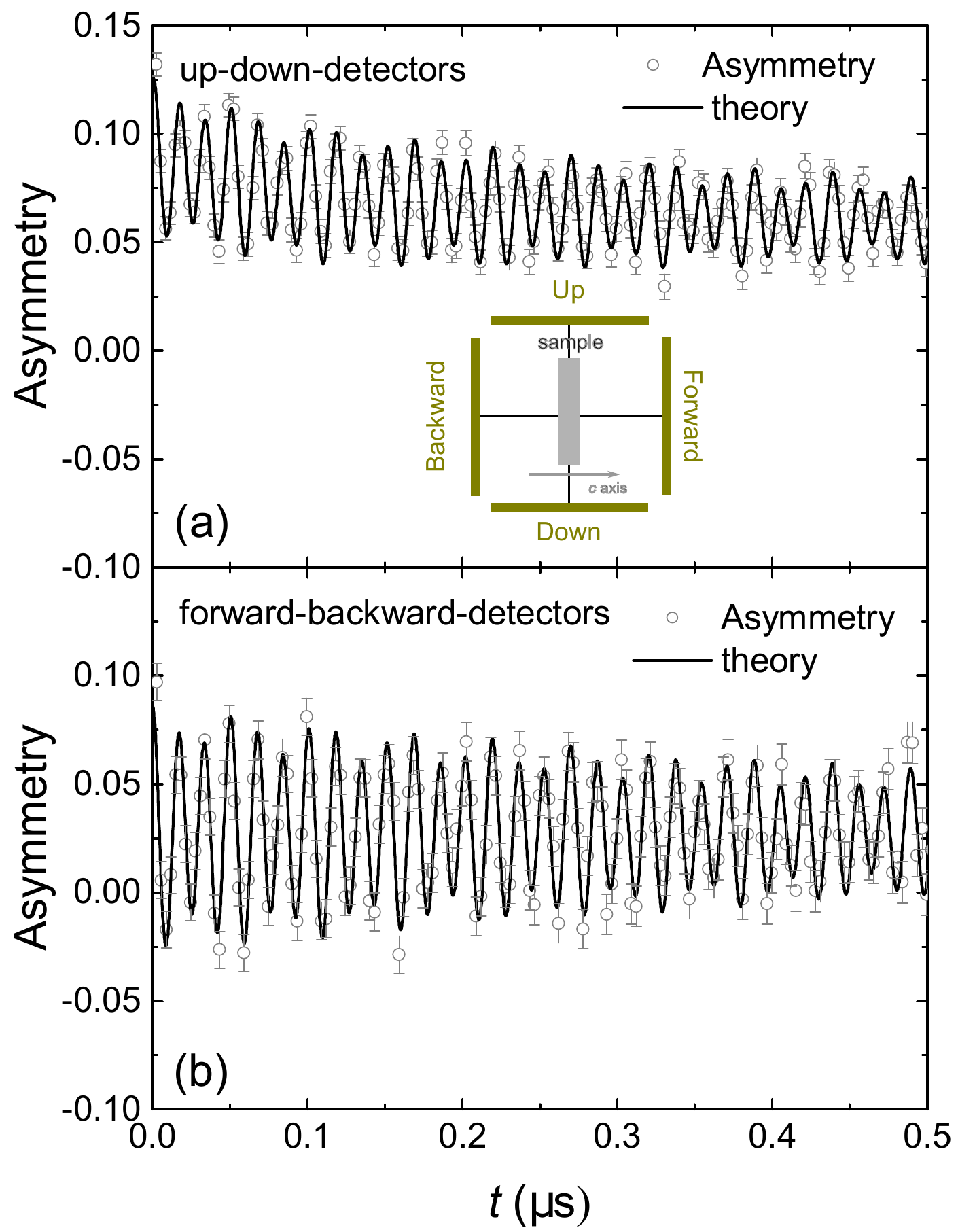} 
\caption{Ambient-pressure zero-field $\mu$SR asymmetry of an oriented single crystal of EuCd$_2$As$_2$ using the GPS spectrometer. The orientation of the sample with respect to the detectors is indicated in the inset of (a). The asymmetry is evaluated from the difference in counts of the up- and down-detectors (a) as well as the forward- and backward-detectors (b).}
\label{fig:muSR-ambient}
\end{figure}
\end{center}

So as to characterize the magnetic order at ambient pressure in more detail, we performed $\mu$SR measurements in zero-field on an oriented single crystal in the GPS spectrometer. Given that the sample is oriented, information on the orientation of the magnetic order can be inferred from differences in counts in the forward- and backward vs. the up- and down-detectors (see Fig.\,\ref{fig:muSR-ambient} for a schematic sketch of the experimental setup). Clear oscillations are seen in the asymmetry evaluated from the difference between the up- and down-detectors (Fig.\,\ref{fig:muSR-ambient} (a)) as well as the forward-backward-detectors (Fig.\,\ref{fig:muSR-ambient} (b)). This implies that the moments are not aligned out-of-plane at ambient pressure, consistent with earlier literature results \cite{Rahn18,Soh19}. The data is best described by taking two muon stopping sites into account and the respective internal fields inferred from these ambient-pressure measurements agree very well with the ones plotted in Fig.\,\ref{fig:internalfield}, which were inferred from the data inside the pressure cell at 0\,GPa.

\begin{center}
\begin{figure}
\includegraphics[width=0.9\columnwidth]{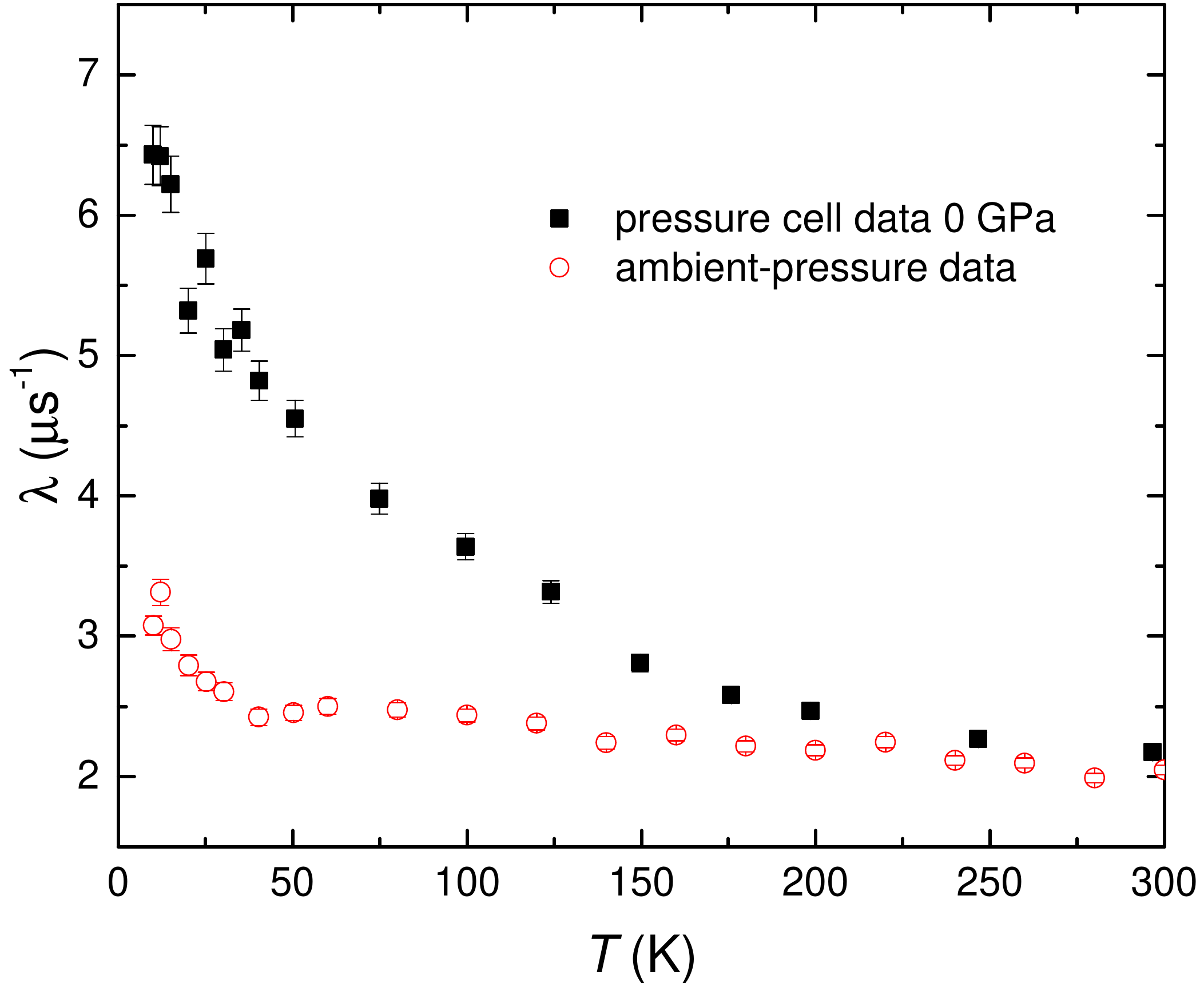} 
\caption{Temperature dependence of the dynamical relaxation rate, $\Lambda$, of EuCd$_2$As$_2$ above the magnetic ordering temperature. Data were taken inside the pressure cell at ambient pressure on a arbitrarily-oriented aggregate of single crystals (black symbols) as well as an oriented single crystal at ambient pressure outside of the pressure cell (red symbols).}
\label{fig:muSR-relaxation-appendix}
\end{figure}
\end{center}

In Fig.\,\ref{fig:muSR-relaxation-appendix}, we show the relaxation rate, $\Lambda$ above the ordering temperature inferred from the GPS experiment on the oriented single crystal and compare it with the data inferred from the randomly-oriented aggregate of single crystals inside the pressure cell at ambient pressure. Whereas the overall magnitude of $\Lambda$ differs between the two experiments, which likely can be attributed to a directional-dependent size of the relaxation, none of the data reveal a clear feature at $\approx$\,100\,K, as previously stated in Ref.\,\cite{Ma19}.

\subsection{Hall effect measurements under pressure}
\label{sec:Hall}

\begin{center}
\begin{figure}
\includegraphics[width=0.9\columnwidth]{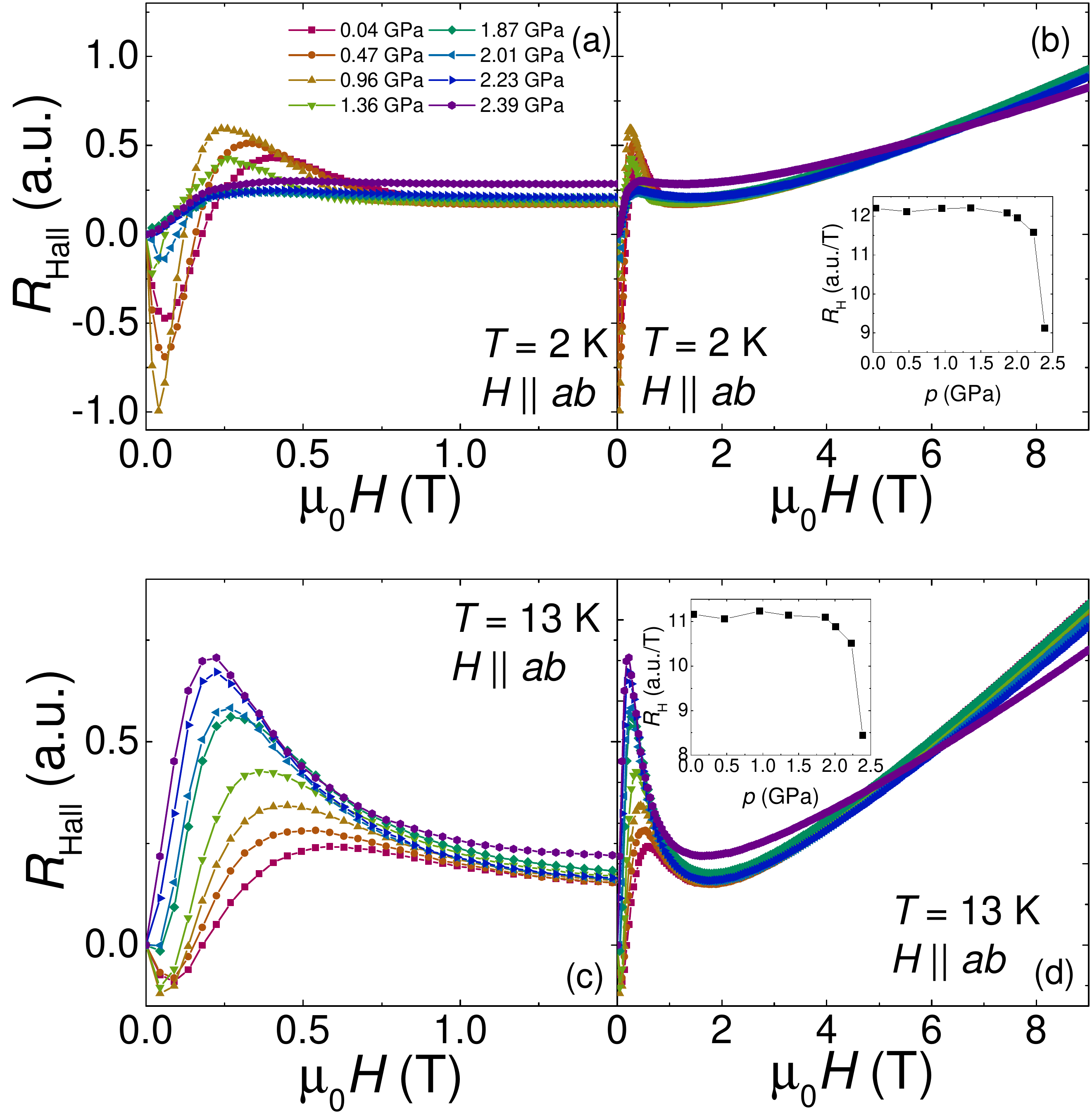} 
\caption{Hall resistance, $R_\textrm{Hall}$, of EuCd$_2$As$_2$ at different pressures at $T\,=\,2\,$K (a,b) and $T\,=\,13\,$K (c,d) in low fields (a,c) and high fields (b,d). Data were taken with magnetic field along in-plane direction. The insets show the high-field slope of the Hall data as a function of pressure at the respective temperatures.}
\label{fig:Hall-MR}
\end{figure}
\end{center}

Hall effect measurements were made by carefully painting two current contacts to cover the two opposite ends of the crystal to ensure as uniform of a current density as possible and two voltage contact on the remaining two side surfaces of the crystal. Current was applied along an in-plane direction, the magnetic field aligned perpendicular to it within the plane and voltage measured along the $c$ axis.

Here, we want to discuss our results of changes of Hall resistance across the critical pressure. In Fig.\,\ref{fig:Hall-MR} we show the Hall resistance, $R_\textrm{Hall}$, as a function of pressure for $T\,=\,2\,$K (a,b) and $T\,=\,13\,$K (c,d), i.e., below and above the magnetic transition temperatures $T_\textrm{N}(p)$ and $T_\textrm{C}(p)$. At 2\,K and low pressures, $R_\textrm{Hall}$ goes through a minimum followed by a maximum in a very narrow field range up to 0.8\,T. The ordinary Hall effect cannot account for this behavior of $R_\textrm{Hall}$ below $\sim\,0.8\,$T as it is almost linear (see Fig.\,\ref{fig:Hall-MR}\,(b) for data up to higher fields). In addition, this field scale matches with the one observed in the MR measurements above as well as magnetization measurements at ambient pressure \cite{Jo20}. Thus, this indicates that the low-field Hall effect data are dominated by the anomalous Hall effect contribution. At high pressures, in the FM state, the pronounced maxima and minima are absent and the anomalous contribution to $R_\textrm{Hall}$ is distinctly smaller, but nonetheless finite. A full disentanglement of the various contributions to the Hall resistance in a magnetic material is unfortunately not possible on the basis of the present data set, given the lack of magnetization data under pressure. 


For sufficiently large fields, $\mu_0H\,\gtrsim\,4\,$T, the $R_\textrm{Hall}$ data at 2\,K and 13\,K are linear in field (see Figs.\,\ref{fig:Hall-MR}\,(b) and (d)), and thus reflects the ordinary Hall contribution. The slope, $R_H$, obtained from linear fits of the Hall data up to maximum field, is plotted as a function for pressure in the insets of Figs.\,\ref{fig:Hall-MR}\,(b) and (d). For each temperature (2\,K and 13\,K), $R_H$ is constant up to $\approx\,2$\,GPa and starts to decrease slightly above. Even though the number of data points at very high pressures, where the decrease becomes pronounced, is limited, our data might indicate a change of charge carrier density across the critical pressure. It will be interesting to investigate the microscopic origin of the change of the magnetic properties with pressure in the future from experimental and theoretical point of view in more detail.

\subsection{Low-temperature lattice parameters of EuCd$_2$As$_2$ at ambient pressure}
\label{sec:xraydiffraction}

\begin{center}
\begin{figure}
\includegraphics[width=0.9\columnwidth]{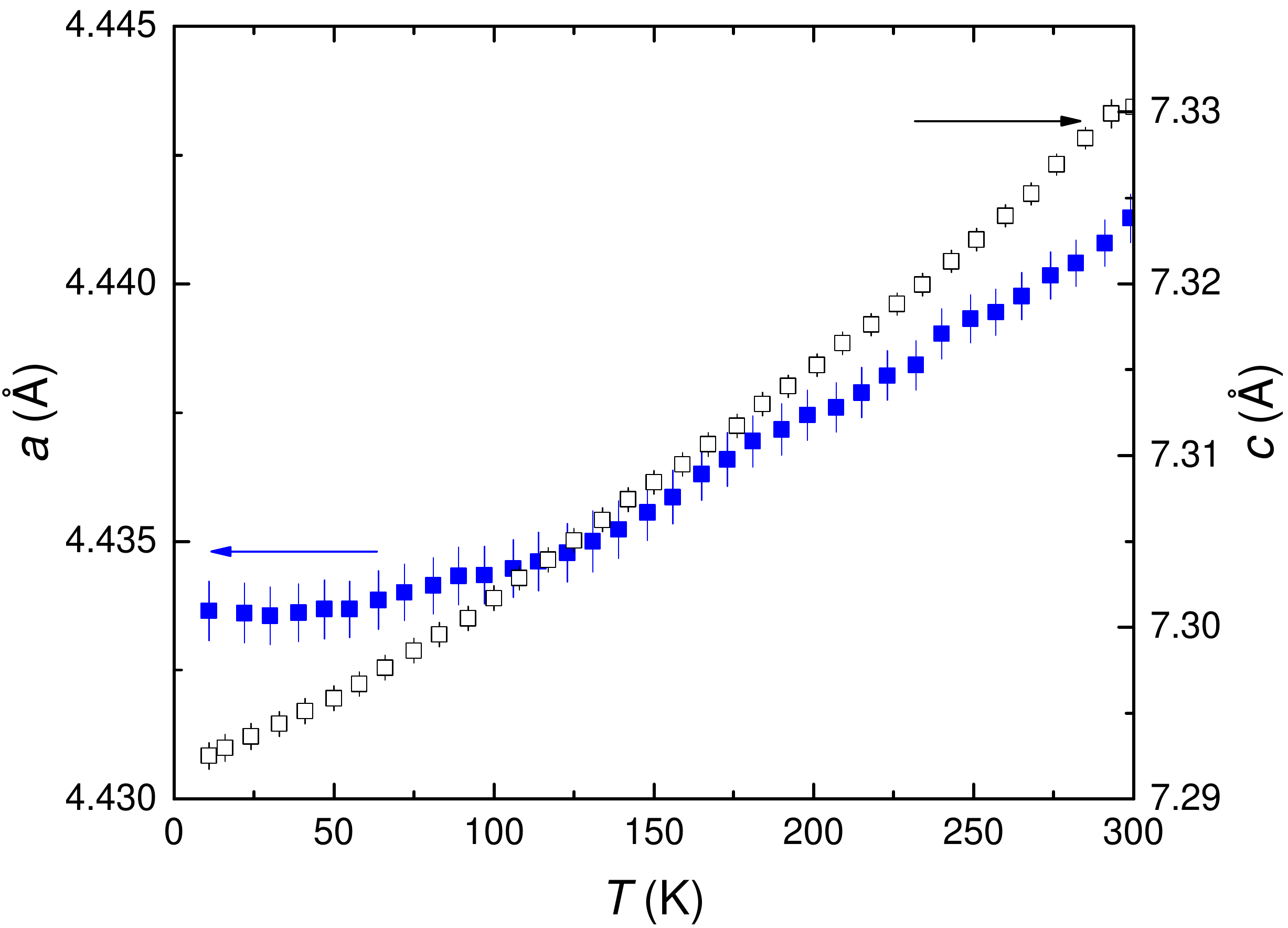} 
\caption{Lattice parameters, $a$ (left axis) and $c$ (right axis) of EuCd$_2$As$_2$ as a function of temperature, $T$, at ambient pressure, determined from x-ray diffraction.}
\label{fig:latticeparameters}
\end{figure}
\end{center}

So as to better compare experimental and theoretical lattice parameters, we determined the lattice parameters of EuCd$_2$As$_2$ experimentally at low temperatures, given that our DFT calculations are performed at zero temperature. Single crystal X-ray diffraction measurements were performed on an in-house four-circle diffractometer using Cu $K_{\alpha 1}$ radiation from a rotating anode X-ray source, using germanium (1\,1\,1) monochromator. A He closed-cycle refrigerator was used for temperature dependence measurements between 11\, and 300\,K. Three Be domes were used as a vacuum shroud, heat shield, and the last innermost dome containing the sample. The innermost dome was filled with a small amount of He gas to improve thermal contact to the sample surface. Measurements were carried out on a single crystal of 0.087\,g attached to a flat copper sample holder that is attached to the cold finger. The mosaicities of the sample were less than 0.04$^\circ$ for both the (0,\,0,\,5) and (3, \,0,\,5) reflections at all measured temperatures. The positions of the reflections were fit using a Lorentzian lineshape, and used to determine the lattice parameters of the sample from 11-300 K, which are shown in Fig.\,\ref{fig:latticeparameters}.

As discussed in the main text, the experimental lattice parameters at 11\,K are within 1\,\% of the calculated values for both the magnetic and the non-magnetic relaxation. Based on previous works \cite{Harl10}, this can be considered as a very good agreement between experiment and theory. 

\subsection{Effect of lattice parameter relaxation in the calculation of energies}\label{app:lattice-parameters-theory}

To justify the validity of our calculated data, we used several slightly different calculation settings. Whereas the different calculations resulted in slightly different estimates of the critical pressures, they always resulted in the AFM$_{ab}$-FM$_{ab}$-FM$_c$ transition sequence with increasing pressure. Thus, the key message of our paper is stable with respect to different calculation settings. In the following, we discuss the spread of $p_c$ values obtained from different calculation procedures.
The order of magnitude for the difference in energy of possible magnetic ground states is already below $1$ meV at ambient pressure, for the MAE even below $0.1$ meV. Therefore, tiny structural changes, caused by different methods of relaxing the unit cell or pseudo potential files, affect the resulting critical pressure $p_c$ measurably.
Independent of all previously shown calculations, we performed a relaxation using the magnetic space groups $C_c 2/m$, corresponding to the AFM$_{ab}$, and $C2/m$, FM$_{ab}$, as starting points for the initial crystal structure. %
Fixing the Wyckoff positions with regard to the magnetic space groups guarantees no loss of symmetry information. Thereby, we also included SOC and a Hubbard $U = 4.4 $ eV. Apart from this, the input parameters are identical to the non-magnetic case.

\begin{center}
\begin{figure}
\includegraphics[width=0.9\columnwidth]{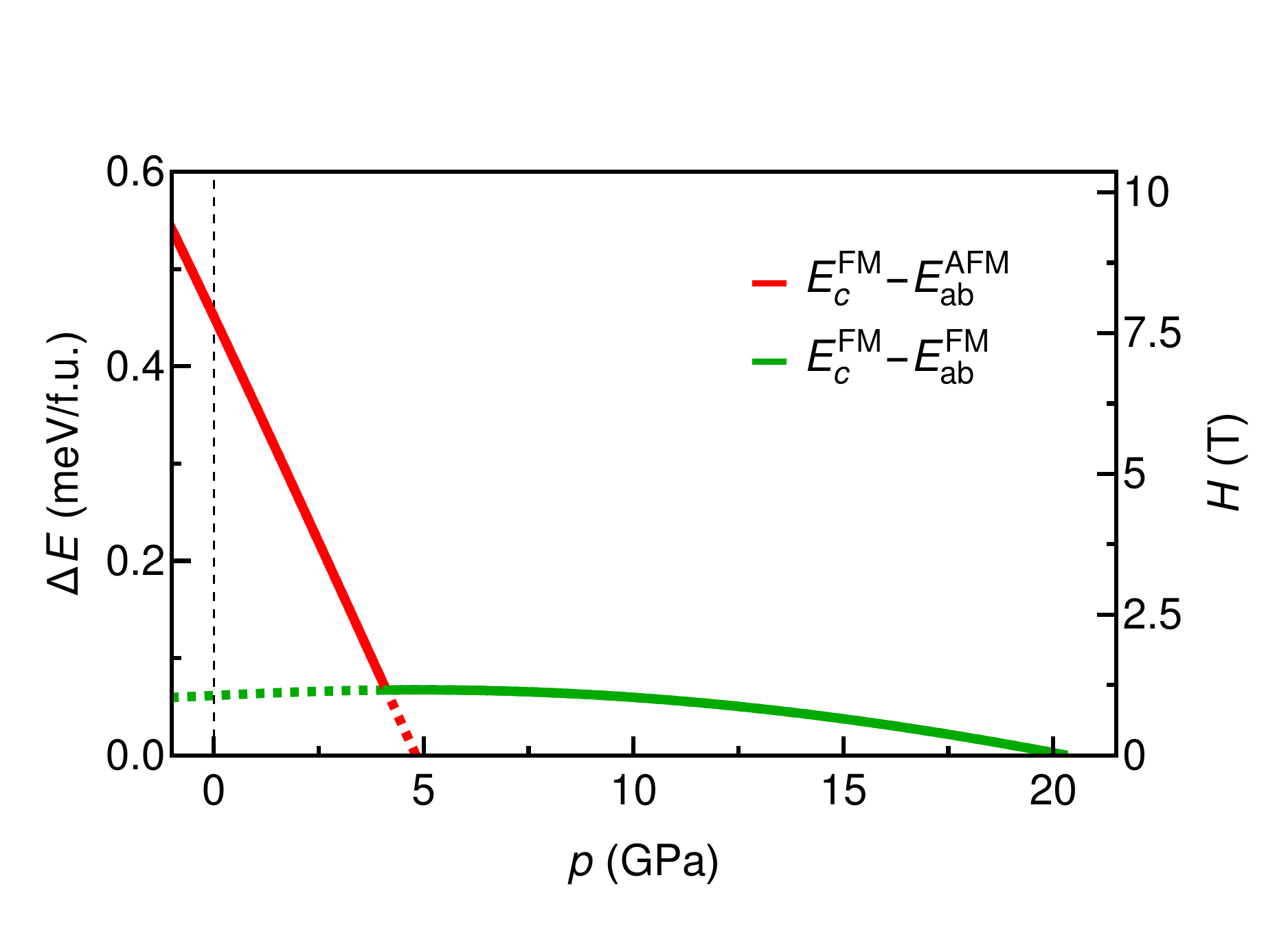} 
\caption{{Energy difference of the ground state compared to FM$_{c}$ order (left axis) and corresponding magnetic field (right axis) vs. pressure. The red curve describes the difference of AFM$_{ab}$ to FM$_{c}$, whereas the green curve describes the difference of FM$_{ab}$ to FM$_{c}$. Compared to the relaxation types in Fig. \ref{fig:EOS} and Fig. \ref{fig:NM-rlx} $p_c$ is slightly increased to $4.2$ GPa and $p'_c$, describing the transition to FM$_c$ ground state, decreased to $20$ GPa. 
}}
\label{fig:AF-rlx}
\end{figure}
\end{center}

Figure \ref{fig:AF-rlx} illustrates the outcome of this method. Similar to Fig.\,\ref{fig:NM-rlx} in the main text, this figure describes the difference of the corresponding ground state energy to the FM$_c$ state as a function of pressure on the left axis and the corresponding magnetic field. The critical pressure $p_c$ for the transition from AFM$_{ab}$ order (red curve) to FM$_{ab}$ (green curve), given by the crossing point of the red and green curve, is shifted to a slightly larger pressure of $4.2$ GPa relative to what is shown in Fig.\,\ref{fig:NM-rlx}. The predicted transition to FM$_c$, occurs at a lower pressure of $p'_c\approx 20$ GPa. 

\bibliographystyle{apsrev}

\begin{thebibliography}{61}
\expandafter\ifx\csname natexlab\endcsname\relax\def\natexlab#1{#1}\fi
\expandafter\ifx\csname bibnamefont\endcsname\relax
  \def\bibnamefont#1{#1}\fi
\expandafter\ifx\csname bibfnamefont\endcsname\relax
  \def\bibfnamefont#1{#1}\fi
\expandafter\ifx\csname citenamefont\endcsname\relax
  \def\citenamefont#1{#1}\fi
\expandafter\ifx\csname url\endcsname\relax
  \def\url#1{\texttt{#1}}\fi
\expandafter\ifx\csname urlprefix\endcsname\relax\def\urlprefix{URL }\fi
\providecommand{\bibinfo}[2]{#2}
\providecommand{\eprint}[2][]{\url{#2}}

\bibitem[{\citenamefont{Wan et~al.}(2011)\citenamefont{Wan, Turner, Vishwanath,
  and Savrasov}}]{Wan11}
\bibinfo{author}{\bibfnamefont{X.}~\bibnamefont{Wan}},
  \bibinfo{author}{\bibfnamefont{A.~M.} \bibnamefont{Turner}},
  \bibinfo{author}{\bibfnamefont{A.}~\bibnamefont{Vishwanath}},
  \bibnamefont{and} \bibinfo{author}{\bibfnamefont{S.~Y.}
  \bibnamefont{Savrasov}}, \bibinfo{journal}{Phys. Rev. B}
  \textbf{\bibinfo{volume}{83}}, \bibinfo{pages}{205101}
  (\bibinfo{year}{2011}).

\bibitem[{\citenamefont{Hasan and Kane}(2010)}]{Hasan10}
\bibinfo{author}{\bibfnamefont{M.~Z.} \bibnamefont{Hasan}} \bibnamefont{and}
  \bibinfo{author}{\bibfnamefont{C.~L.} \bibnamefont{Kane}},
  \bibinfo{journal}{Rev. Mod. Phys.} \textbf{\bibinfo{volume}{82}},
  \bibinfo{pages}{3045} (\bibinfo{year}{2010}).

\bibitem[{\citenamefont{Yan and Felser}(2017)}]{Yan17}
\bibinfo{author}{\bibfnamefont{B.}~\bibnamefont{Yan}} \bibnamefont{and}
  \bibinfo{author}{\bibfnamefont{C.}~\bibnamefont{Felser}},
  \bibinfo{journal}{Annual Review of Condensed Matter Physics}
  \textbf{\bibinfo{volume}{8}}, \bibinfo{pages}{337} (\bibinfo{year}{2017}).

\bibitem[{\citenamefont{Burkov}(2016)}]{Burkov16}
\bibinfo{author}{\bibfnamefont{A.}~\bibnamefont{Burkov}},
  \bibinfo{journal}{Nature Materials} \textbf{\bibinfo{volume}{15}},
  \bibinfo{pages}{1146} (\bibinfo{year}{2016}).

\bibitem[{\citenamefont{Zhang et~al.}(2019)\citenamefont{Zhang, Burkov, Martin,
  and Heinonen}}]{Zhang19}
\bibinfo{author}{\bibfnamefont{S.~S.-L.} \bibnamefont{Zhang}},
  \bibinfo{author}{\bibfnamefont{A.~A.} \bibnamefont{Burkov}},
  \bibinfo{author}{\bibfnamefont{I.}~\bibnamefont{Martin}}, \bibnamefont{and}
  \bibinfo{author}{\bibfnamefont{O.~G.} \bibnamefont{Heinonen}},
  \bibinfo{journal}{Phys. Rev. Lett.} \textbf{\bibinfo{volume}{123}},
  \bibinfo{pages}{187201} (\bibinfo{year}{2019}).

\bibitem[{\citenamefont{Šmejkal et~al.}(2017)\citenamefont{Šmejkal,
  Jungwirth, and Sinova}}]{Smejkal17}
\bibinfo{author}{\bibfnamefont{L.}~\bibnamefont{Šmejkal}},
  \bibinfo{author}{\bibfnamefont{T.}~\bibnamefont{Jungwirth}},
  \bibnamefont{and} \bibinfo{author}{\bibfnamefont{J.}~\bibnamefont{Sinova}},
  \bibinfo{journal}{physica status solidi (RRL)} \textbf{\bibinfo{volume}{11}},
  \bibinfo{pages}{1700044} (\bibinfo{year}{2017}).

\bibitem[{\citenamefont{Armitage et~al.}(2018)\citenamefont{Armitage, Mele, and
  Vishwanath}}]{Armitage18}
\bibinfo{author}{\bibfnamefont{N.~P.} \bibnamefont{Armitage}},
  \bibinfo{author}{\bibfnamefont{E.~J.} \bibnamefont{Mele}}, \bibnamefont{and}
  \bibinfo{author}{\bibfnamefont{A.}~\bibnamefont{Vishwanath}},
  \bibinfo{journal}{Rev. Mod. Phys.} \textbf{\bibinfo{volume}{90}},
  \bibinfo{pages}{015001} (\bibinfo{year}{2018}).

\bibitem[{\citenamefont{Burkov and Balents}(2011)}]{Burkov11}
\bibinfo{author}{\bibfnamefont{A.~A.} \bibnamefont{Burkov}} \bibnamefont{and}
  \bibinfo{author}{\bibfnamefont{L.}~\bibnamefont{Balents}},
  \bibinfo{journal}{Phys. Rev. Lett.} \textbf{\bibinfo{volume}{107}},
  \bibinfo{pages}{127205} (\bibinfo{year}{2011}).

\bibitem[{\citenamefont{Xu et~al.}(2011)\citenamefont{Xu, Weng, Wang, Dai, and
  Fang}}]{Xu11}
\bibinfo{author}{\bibfnamefont{G.}~\bibnamefont{Xu}},
  \bibinfo{author}{\bibfnamefont{H.}~\bibnamefont{Weng}},
  \bibinfo{author}{\bibfnamefont{Z.}~\bibnamefont{Wang}},
  \bibinfo{author}{\bibfnamefont{X.}~\bibnamefont{Dai}}, \bibnamefont{and}
  \bibinfo{author}{\bibfnamefont{Z.}~\bibnamefont{Fang}},
  \bibinfo{journal}{Phys. Rev. Lett.} \textbf{\bibinfo{volume}{107}},
  \bibinfo{pages}{186806} (\bibinfo{year}{2011}).

\bibitem[{\citenamefont{Son and Spivak}(2013)}]{Son13}
\bibinfo{author}{\bibfnamefont{D.~T.} \bibnamefont{Son}} \bibnamefont{and}
  \bibinfo{author}{\bibfnamefont{B.~Z.} \bibnamefont{Spivak}},
  \bibinfo{journal}{Phys. Rev. B} \textbf{\bibinfo{volume}{88}},
  \bibinfo{pages}{104412} (\bibinfo{year}{2013}).

\bibitem[{\citenamefont{Parameswaran et~al.}(2014)\citenamefont{Parameswaran,
  Grover, Abanin, Pesin, and Vishwanath}}]{Parameswaran14}
\bibinfo{author}{\bibfnamefont{S.~A.} \bibnamefont{Parameswaran}},
  \bibinfo{author}{\bibfnamefont{T.}~\bibnamefont{Grover}},
  \bibinfo{author}{\bibfnamefont{D.~A.} \bibnamefont{Abanin}},
  \bibinfo{author}{\bibfnamefont{D.~A.} \bibnamefont{Pesin}}, \bibnamefont{and}
  \bibinfo{author}{\bibfnamefont{A.}~\bibnamefont{Vishwanath}},
  \bibinfo{journal}{Phys. Rev. X} \textbf{\bibinfo{volume}{4}},
  \bibinfo{pages}{031035} (\bibinfo{year}{2014}).

\bibitem[{\citenamefont{Liu and Vanderbilt}(2014)}]{Liu14}
\bibinfo{author}{\bibfnamefont{J.}~\bibnamefont{Liu}} \bibnamefont{and}
  \bibinfo{author}{\bibfnamefont{D.}~\bibnamefont{Vanderbilt}},
  \bibinfo{journal}{Phys. Rev. B} \textbf{\bibinfo{volume}{90}},
  \bibinfo{pages}{155316} (\bibinfo{year}{2014}).

\bibitem[{\citenamefont{Heinsdorf et~al.}(2021)\citenamefont{Heinsdorf,
  Christensen, Iraola, Zhang, Yang, Birol, Batista, Valent{\'\i}, and
  Fernandes}}]{heinsdorf2021}
\bibinfo{author}{\bibfnamefont{N.}~\bibnamefont{Heinsdorf}},
  \bibinfo{author}{\bibfnamefont{M.~H.} \bibnamefont{Christensen}},
  \bibinfo{author}{\bibfnamefont{M.}~\bibnamefont{Iraola}},
  \bibinfo{author}{\bibfnamefont{S.}~\bibnamefont{Zhang}},
  \bibinfo{author}{\bibfnamefont{F.}~\bibnamefont{Yang}},
  \bibinfo{author}{\bibfnamefont{T.}~\bibnamefont{Birol}},
  \bibinfo{author}{\bibfnamefont{C.~D.} \bibnamefont{Batista}},
  \bibinfo{author}{\bibfnamefont{R.}~\bibnamefont{Valent{\'\i}}},
  \bibnamefont{and} \bibinfo{author}{\bibfnamefont{R.~M.}
  \bibnamefont{Fernandes}}, \bibinfo{journal}{arXiv preprint arXiv:2101.05301}
  (\bibinfo{year}{2021}).

\bibitem[{\citenamefont{Chang et~al.}(2018)\citenamefont{Chang, Singh, Xu,
  Bian, Huang, Hsu, Belopolski, Alidoust, Sanchez, Zheng et~al.}}]{Chang18}
\bibinfo{author}{\bibfnamefont{G.}~\bibnamefont{Chang}},
  \bibinfo{author}{\bibfnamefont{B.}~\bibnamefont{Singh}},
  \bibinfo{author}{\bibfnamefont{S.-Y.} \bibnamefont{Xu}},
  \bibinfo{author}{\bibfnamefont{G.}~\bibnamefont{Bian}},
  \bibinfo{author}{\bibfnamefont{S.-M.} \bibnamefont{Huang}},
  \bibinfo{author}{\bibfnamefont{C.-H.} \bibnamefont{Hsu}},
  \bibinfo{author}{\bibfnamefont{I.}~\bibnamefont{Belopolski}},
  \bibinfo{author}{\bibfnamefont{N.}~\bibnamefont{Alidoust}},
  \bibinfo{author}{\bibfnamefont{D.~S.} \bibnamefont{Sanchez}},
  \bibinfo{author}{\bibfnamefont{H.}~\bibnamefont{Zheng}},
  \bibnamefont{et~al.}, \bibinfo{journal}{Phys. Rev. B}
  \textbf{\bibinfo{volume}{97}}, \bibinfo{pages}{041104}
  (\bibinfo{year}{2018}).

\bibitem[{\citenamefont{Arnold et~al.}(2016)\citenamefont{Arnold, Shekhar, Wu,
  Sun, dos Reis, Kumar, Naumann, Ajeesh, Schmidt, Grushin et~al.}}]{Arnold16}
\bibinfo{author}{\bibfnamefont{F.}~\bibnamefont{Arnold}},
  \bibinfo{author}{\bibfnamefont{C.}~\bibnamefont{Shekhar}},
  \bibinfo{author}{\bibfnamefont{S.-C.} \bibnamefont{Wu}},
  \bibinfo{author}{\bibfnamefont{Y.}~\bibnamefont{Sun}},
  \bibinfo{author}{\bibfnamefont{R.~D.} \bibnamefont{dos Reis}},
  \bibinfo{author}{\bibfnamefont{N.}~\bibnamefont{Kumar}},
  \bibinfo{author}{\bibfnamefont{M.}~\bibnamefont{Naumann}},
  \bibinfo{author}{\bibfnamefont{M.~O.} \bibnamefont{Ajeesh}},
  \bibinfo{author}{\bibfnamefont{M.}~\bibnamefont{Schmidt}},
  \bibinfo{author}{\bibfnamefont{A.~G.} \bibnamefont{Grushin}},
  \bibnamefont{et~al.}, \bibinfo{journal}{Nat. Commun.}
  \textbf{\bibinfo{volume}{7}}, \bibinfo{pages}{11615} (\bibinfo{year}{2016}).

\bibitem[{\citenamefont{Weng et~al.}(2015)\citenamefont{Weng, Fang, Fang,
  Bernevig, and Dai}}]{Weng15}
\bibinfo{author}{\bibfnamefont{H.}~\bibnamefont{Weng}},
  \bibinfo{author}{\bibfnamefont{C.}~\bibnamefont{Fang}},
  \bibinfo{author}{\bibfnamefont{Z.}~\bibnamefont{Fang}},
  \bibinfo{author}{\bibfnamefont{B.~A.} \bibnamefont{Bernevig}},
  \bibnamefont{and} \bibinfo{author}{\bibfnamefont{X.}~\bibnamefont{Dai}},
  \bibinfo{journal}{Phys. Rev. X} \textbf{\bibinfo{volume}{5}},
  \bibinfo{pages}{011029} (\bibinfo{year}{2015}).

\bibitem[{\citenamefont{Huang et~al.}(2015)\citenamefont{Huang, Xu, Belopolski,
  Lee, Chang, Wang, Alidoust, Bian, Neupane, Zhang et~al.}}]{Huang15b}
\bibinfo{author}{\bibfnamefont{S.-M.} \bibnamefont{Huang}},
  \bibinfo{author}{\bibfnamefont{S.-Y.} \bibnamefont{Xu}},
  \bibinfo{author}{\bibfnamefont{I.}~\bibnamefont{Belopolski}},
  \bibinfo{author}{\bibfnamefont{C.-C.} \bibnamefont{Lee}},
  \bibinfo{author}{\bibfnamefont{G.}~\bibnamefont{Chang}},
  \bibinfo{author}{\bibfnamefont{B.}~\bibnamefont{Wang}},
  \bibinfo{author}{\bibfnamefont{N.}~\bibnamefont{Alidoust}},
  \bibinfo{author}{\bibfnamefont{G.}~\bibnamefont{Bian}},
  \bibinfo{author}{\bibfnamefont{M.}~\bibnamefont{Neupane}},
  \bibinfo{author}{\bibfnamefont{C.}~\bibnamefont{Zhang}},
  \bibnamefont{et~al.}, \bibinfo{journal}{Nat. Commun.}
  (\bibinfo{year}{2015}).

\bibitem[{\citenamefont{Borisenko et~al.}(2019)\citenamefont{Borisenko,
  Evtushinsky, Gibson, Yaresko, Koepernik, Kim, Ali, van~den Brink, Hoesch,
  Fedorov et~al.}}]{Borisenko19}
\bibinfo{author}{\bibfnamefont{S.}~\bibnamefont{Borisenko}},
  \bibinfo{author}{\bibfnamefont{D.}~\bibnamefont{Evtushinsky}},
  \bibinfo{author}{\bibfnamefont{Q.}~\bibnamefont{Gibson}},
  \bibinfo{author}{\bibfnamefont{A.}~\bibnamefont{Yaresko}},
  \bibinfo{author}{\bibfnamefont{K.}~\bibnamefont{Koepernik}},
  \bibinfo{author}{\bibfnamefont{T.}~\bibnamefont{Kim}},
  \bibinfo{author}{\bibfnamefont{M.}~\bibnamefont{Ali}},
  \bibinfo{author}{\bibfnamefont{J.}~\bibnamefont{van~den Brink}},
  \bibinfo{author}{\bibfnamefont{M.}~\bibnamefont{Hoesch}},
  \bibinfo{author}{\bibfnamefont{A.}~\bibnamefont{Fedorov}},
  \bibnamefont{et~al.}, \bibinfo{journal}{Nat. Commun.}
  \textbf{\bibinfo{volume}{10}}, \bibinfo{pages}{3424} (\bibinfo{year}{2019}).

\bibitem[{\citenamefont{Liu et~al.}(2018)\citenamefont{Liu, Sun, Kumar,
  Muechler, Sun, Jiao, Yang, Liu, Liang, Xu et~al.}}]{Liu18}
\bibinfo{author}{\bibfnamefont{E.}~\bibnamefont{Liu}},
  \bibinfo{author}{\bibfnamefont{Y.}~\bibnamefont{Sun}},
  \bibinfo{author}{\bibfnamefont{N.}~\bibnamefont{Kumar}},
  \bibinfo{author}{\bibfnamefont{L.}~\bibnamefont{Muechler}},
  \bibinfo{author}{\bibfnamefont{A.}~\bibnamefont{Sun}},
  \bibinfo{author}{\bibfnamefont{L.}~\bibnamefont{Jiao}},
  \bibinfo{author}{\bibfnamefont{S.-Y.} \bibnamefont{Yang}},
  \bibinfo{author}{\bibfnamefont{D.}~\bibnamefont{Liu}},
  \bibinfo{author}{\bibfnamefont{A.}~\bibnamefont{Liang}},
  \bibinfo{author}{\bibfnamefont{Q.}~\bibnamefont{Xu}}, \bibnamefont{et~al.},
  \bibinfo{journal}{Nat. Phys.} \textbf{\bibinfo{volume}{14}}
  (\bibinfo{year}{2018}).

\bibitem[{\citenamefont{Ye et~al.}(2018)\citenamefont{Ye, Kang, Liu, von Cube,
  Wicker, Suzuki, Jozwiak, Bostwick, Rotenberg, Bell et~al.}}]{Ye18}
\bibinfo{author}{\bibfnamefont{L.}~\bibnamefont{Ye}},
  \bibinfo{author}{\bibfnamefont{M.}~\bibnamefont{Kang}},
  \bibinfo{author}{\bibfnamefont{J.}~\bibnamefont{Liu}},
  \bibinfo{author}{\bibfnamefont{F.}~\bibnamefont{von Cube}},
  \bibinfo{author}{\bibfnamefont{C.~R.} \bibnamefont{Wicker}},
  \bibinfo{author}{\bibfnamefont{T.}~\bibnamefont{Suzuki}},
  \bibinfo{author}{\bibfnamefont{C.}~\bibnamefont{Jozwiak}},
  \bibinfo{author}{\bibfnamefont{A.}~\bibnamefont{Bostwick}},
  \bibinfo{author}{\bibfnamefont{E.}~\bibnamefont{Rotenberg}},
  \bibinfo{author}{\bibfnamefont{D.~C.} \bibnamefont{Bell}},
  \bibnamefont{et~al.}, \bibinfo{journal}{Nature}
  \textbf{\bibinfo{volume}{555}} (\bibinfo{year}{2018}).

\bibitem[{\citenamefont{Kang et~al.}(2020)\citenamefont{Kang, Ye, Fang, You,
  Levitan, Han, Facio, Jozwiak, Bostwick, Rotenberg et~al.}}]{Kang20}
\bibinfo{author}{\bibfnamefont{M.}~\bibnamefont{Kang}},
  \bibinfo{author}{\bibfnamefont{L.}~\bibnamefont{Ye}},
  \bibinfo{author}{\bibfnamefont{S.}~\bibnamefont{Fang}},
  \bibinfo{author}{\bibfnamefont{J.-S.} \bibnamefont{You}},
  \bibinfo{author}{\bibfnamefont{A.}~\bibnamefont{Levitan}},
  \bibinfo{author}{\bibfnamefont{M.}~\bibnamefont{Han}},
  \bibinfo{author}{\bibfnamefont{J.~I.} \bibnamefont{Facio}},
  \bibinfo{author}{\bibfnamefont{C.}~\bibnamefont{Jozwiak}},
  \bibinfo{author}{\bibfnamefont{A.}~\bibnamefont{Bostwick}},
  \bibinfo{author}{\bibfnamefont{E.}~\bibnamefont{Rotenberg}},
  \bibnamefont{et~al.}, \bibinfo{journal}{Nat. Mat.}  (\bibinfo{year}{2020}).

\bibitem[{\citenamefont{Guguchia et~al.}(2020)\citenamefont{Guguchia, Verezhak,
  Gawryluk, Tsirkin, Yin, Belopolski, Zhou, Simutis, Zhang, Cochran
  et~al.}}]{Guguchia20}
\bibinfo{author}{\bibfnamefont{Z.}~\bibnamefont{Guguchia}},
  \bibinfo{author}{\bibfnamefont{J.~A.~T.} \bibnamefont{Verezhak}},
  \bibinfo{author}{\bibfnamefont{D.~J.} \bibnamefont{Gawryluk}},
  \bibinfo{author}{\bibfnamefont{S.~S.} \bibnamefont{Tsirkin}},
  \bibinfo{author}{\bibfnamefont{J.-X.} \bibnamefont{Yin}},
  \bibinfo{author}{\bibfnamefont{I.}~\bibnamefont{Belopolski}},
  \bibinfo{author}{\bibfnamefont{H.}~\bibnamefont{Zhou}},
  \bibinfo{author}{\bibfnamefont{G.}~\bibnamefont{Simutis}},
  \bibinfo{author}{\bibfnamefont{S.-S.} \bibnamefont{Zhang}},
  \bibinfo{author}{\bibfnamefont{T.~A.} \bibnamefont{Cochran}},
  \bibnamefont{et~al.}, \bibinfo{journal}{Nat. Commun.}
  \textbf{\bibinfo{volume}{11}} (\bibinfo{year}{2020}).

\bibitem[{\citenamefont{Destraz et~al.}(2020)\citenamefont{Destraz, Das,
  Tsirkin, Xu, Neupert, Chang, Schilling, Grushin, Kohlbrecher, Keller
  et~al.}}]{Destraz20}
\bibinfo{author}{\bibfnamefont{D.}~\bibnamefont{Destraz}},
  \bibinfo{author}{\bibfnamefont{L.}~\bibnamefont{Das}},
  \bibinfo{author}{\bibfnamefont{S.~S.} \bibnamefont{Tsirkin}},
  \bibinfo{author}{\bibfnamefont{Y.}~\bibnamefont{Xu}},
  \bibinfo{author}{\bibfnamefont{T.}~\bibnamefont{Neupert}},
  \bibinfo{author}{\bibfnamefont{J.}~\bibnamefont{Chang}},
  \bibinfo{author}{\bibfnamefont{A.}~\bibnamefont{Schilling}},
  \bibinfo{author}{\bibfnamefont{A.~G.} \bibnamefont{Grushin}},
  \bibinfo{author}{\bibfnamefont{J.}~\bibnamefont{Kohlbrecher}},
  \bibinfo{author}{\bibfnamefont{L.}~\bibnamefont{Keller}},
  \bibnamefont{et~al.}, \bibinfo{journal}{npj Quantum Materials}
  \textbf{\bibinfo{volume}{5}} (\bibinfo{year}{2020}).

\bibitem[{\citenamefont{Hua et~al.}(2018)\citenamefont{Hua, Nie, Song, Yu, Xu,
  and Yao}}]{Hua18}
\bibinfo{author}{\bibfnamefont{G.}~\bibnamefont{Hua}},
  \bibinfo{author}{\bibfnamefont{S.}~\bibnamefont{Nie}},
  \bibinfo{author}{\bibfnamefont{Z.}~\bibnamefont{Song}},
  \bibinfo{author}{\bibfnamefont{R.}~\bibnamefont{Yu}},
  \bibinfo{author}{\bibfnamefont{G.}~\bibnamefont{Xu}}, \bibnamefont{and}
  \bibinfo{author}{\bibfnamefont{K.}~\bibnamefont{Yao}},
  \bibinfo{journal}{Phys. Rev. B} \textbf{\bibinfo{volume}{98}},
  \bibinfo{pages}{201116} (\bibinfo{year}{2018}).

\bibitem[{\citenamefont{Wang et~al.}(2019)\citenamefont{Wang, Jo, Kuthanazhi,
  Wu, McQueeney, Kaminski, and Canfield}}]{Wang19}
\bibinfo{author}{\bibfnamefont{L.-L.} \bibnamefont{Wang}},
  \bibinfo{author}{\bibfnamefont{N.~H.} \bibnamefont{Jo}},
  \bibinfo{author}{\bibfnamefont{B.}~\bibnamefont{Kuthanazhi}},
  \bibinfo{author}{\bibfnamefont{Y.}~\bibnamefont{Wu}},
  \bibinfo{author}{\bibfnamefont{R.~J.} \bibnamefont{McQueeney}},
  \bibinfo{author}{\bibfnamefont{A.}~\bibnamefont{Kaminski}}, \bibnamefont{and}
  \bibinfo{author}{\bibfnamefont{P.~C.} \bibnamefont{Canfield}},
  \bibinfo{journal}{Phys. Rev. B} \textbf{\bibinfo{volume}{99}},
  \bibinfo{pages}{245147} (\bibinfo{year}{2019}).

\bibitem[{\citenamefont{Artmann et~al.}(1996)\citenamefont{Artmann, Mewis,
  Roepke, and Michels}}]{Artmann96}
\bibinfo{author}{\bibfnamefont{A.}~\bibnamefont{Artmann}},
  \bibinfo{author}{\bibfnamefont{A.}~\bibnamefont{Mewis}},
  \bibinfo{author}{\bibfnamefont{M.}~\bibnamefont{Roepke}}, \bibnamefont{and}
  \bibinfo{author}{\bibfnamefont{G.}~\bibnamefont{Michels}},
  \bibinfo{journal}{Zeitschrift für anorganische und allgemeine Chemie}
  \textbf{\bibinfo{volume}{622}}, \bibinfo{pages}{679} (\bibinfo{year}{1996}).

\bibitem[{\citenamefont{Wang et~al.}(2016)\citenamefont{Wang, Wu, Shi, and
  Wang}}]{Wang16}
\bibinfo{author}{\bibfnamefont{H.~P.} \bibnamefont{Wang}},
  \bibinfo{author}{\bibfnamefont{D.~S.} \bibnamefont{Wu}},
  \bibinfo{author}{\bibfnamefont{Y.~G.} \bibnamefont{Shi}}, \bibnamefont{and}
  \bibinfo{author}{\bibfnamefont{N.~L.} \bibnamefont{Wang}},
  \bibinfo{journal}{Phys. Rev. B} \textbf{\bibinfo{volume}{94}},
  \bibinfo{pages}{045112} (\bibinfo{year}{2016}).

\bibitem[{\citenamefont{Rahn et~al.}(2018)\citenamefont{Rahn, Soh, Francoual,
  Veiga, Strempfer, Mardegan, Yan, Guo, Shi, and Boothroyd}}]{Rahn18}
\bibinfo{author}{\bibfnamefont{M.~C.} \bibnamefont{Rahn}},
  \bibinfo{author}{\bibfnamefont{J.-R.} \bibnamefont{Soh}},
  \bibinfo{author}{\bibfnamefont{S.}~\bibnamefont{Francoual}},
  \bibinfo{author}{\bibfnamefont{L.~S.~I.} \bibnamefont{Veiga}},
  \bibinfo{author}{\bibfnamefont{J.}~\bibnamefont{Strempfer}},
  \bibinfo{author}{\bibfnamefont{J.}~\bibnamefont{Mardegan}},
  \bibinfo{author}{\bibfnamefont{D.~Y.} \bibnamefont{Yan}},
  \bibinfo{author}{\bibfnamefont{Y.~F.} \bibnamefont{Guo}},
  \bibinfo{author}{\bibfnamefont{Y.~G.} \bibnamefont{Shi}}, \bibnamefont{and}
  \bibinfo{author}{\bibfnamefont{A.~T.} \bibnamefont{Boothroyd}},
  \bibinfo{journal}{Phys. Rev. B} \textbf{\bibinfo{volume}{97}},
  \bibinfo{pages}{214422} (\bibinfo{year}{2018}).

\bibitem[{\citenamefont{Soh et~al.}(2019)\citenamefont{Soh, de~Juan, Vergniory,
  Schr\"oter, Rahn, Yan, Jiang, Bristow, Reiss, Blandy et~al.}}]{Soh19}
\bibinfo{author}{\bibfnamefont{J.-R.} \bibnamefont{Soh}},
  \bibinfo{author}{\bibfnamefont{F.}~\bibnamefont{de~Juan}},
  \bibinfo{author}{\bibfnamefont{M.~G.} \bibnamefont{Vergniory}},
  \bibinfo{author}{\bibfnamefont{N.~B.~M.} \bibnamefont{Schr\"oter}},
  \bibinfo{author}{\bibfnamefont{M.~C.} \bibnamefont{Rahn}},
  \bibinfo{author}{\bibfnamefont{D.~Y.} \bibnamefont{Yan}},
  \bibinfo{author}{\bibfnamefont{J.}~\bibnamefont{Jiang}},
  \bibinfo{author}{\bibfnamefont{M.}~\bibnamefont{Bristow}},
  \bibinfo{author}{\bibfnamefont{P.}~\bibnamefont{Reiss}},
  \bibinfo{author}{\bibfnamefont{J.~N.} \bibnamefont{Blandy}},
  \bibnamefont{et~al.}, \bibinfo{journal}{Phys. Rev. B}
  \textbf{\bibinfo{volume}{100}}, \bibinfo{pages}{201102}
  (\bibinfo{year}{2019}).

\bibitem[{\citenamefont{Ma et~al.}(2019)\citenamefont{Ma, Nie, Yi, Jandke,
  Shang, Yao, Naamneh, Yan, Sun, Chikina et~al.}}]{Ma19}
\bibinfo{author}{\bibfnamefont{J.-Z.} \bibnamefont{Ma}},
  \bibinfo{author}{\bibfnamefont{S.~M.} \bibnamefont{Nie}},
  \bibinfo{author}{\bibfnamefont{C.~J.} \bibnamefont{Yi}},
  \bibinfo{author}{\bibfnamefont{J.}~\bibnamefont{Jandke}},
  \bibinfo{author}{\bibfnamefont{T.}~\bibnamefont{Shang}},
  \bibinfo{author}{\bibfnamefont{M.~Y.} \bibnamefont{Yao}},
  \bibinfo{author}{\bibfnamefont{M.}~\bibnamefont{Naamneh}},
  \bibinfo{author}{\bibfnamefont{L.~Q.} \bibnamefont{Yan}},
  \bibinfo{author}{\bibfnamefont{Y.}~\bibnamefont{Sun}},
  \bibinfo{author}{\bibfnamefont{A.}~\bibnamefont{Chikina}},
  \bibnamefont{et~al.}, \bibinfo{journal}{Science Advances}
  \textbf{\bibinfo{volume}{5}}, \bibinfo{pages}{eaaw4718}
  (\bibinfo{year}{2019}).

\bibitem[{\citenamefont{Niu et~al.}(2019)\citenamefont{Niu, Mao, Hu, Huang, and
  Dai}}]{Niu19}
\bibinfo{author}{\bibfnamefont{C.}~\bibnamefont{Niu}},
  \bibinfo{author}{\bibfnamefont{N.}~\bibnamefont{Mao}},
  \bibinfo{author}{\bibfnamefont{X.}~\bibnamefont{Hu}},
  \bibinfo{author}{\bibfnamefont{B.}~\bibnamefont{Huang}}, \bibnamefont{and}
  \bibinfo{author}{\bibfnamefont{Y.}~\bibnamefont{Dai}},
  \bibinfo{journal}{Phys. Rev. B} \textbf{\bibinfo{volume}{99}},
  \bibinfo{pages}{235119} (\bibinfo{year}{2019}).

\bibitem[{\citenamefont{Xu et~al.}(2021)\citenamefont{Xu, Das, Ma, Yi, Nie,
  Shi, Tiwari, Tsirkin, Neupert, Medarde et~al.}}]{Xu21}
\bibinfo{author}{\bibfnamefont{Y.}~\bibnamefont{Xu}},
  \bibinfo{author}{\bibfnamefont{L.}~\bibnamefont{Das}},
  \bibinfo{author}{\bibfnamefont{J.~Z.} \bibnamefont{Ma}},
  \bibinfo{author}{\bibfnamefont{C.~J.} \bibnamefont{Yi}},
  \bibinfo{author}{\bibfnamefont{S.~M.} \bibnamefont{Nie}},
  \bibinfo{author}{\bibfnamefont{Y.~G.} \bibnamefont{Shi}},
  \bibinfo{author}{\bibfnamefont{A.}~\bibnamefont{Tiwari}},
  \bibinfo{author}{\bibfnamefont{S.~S.} \bibnamefont{Tsirkin}},
  \bibinfo{author}{\bibfnamefont{T.}~\bibnamefont{Neupert}},
  \bibinfo{author}{\bibfnamefont{M.}~\bibnamefont{Medarde}},
  \bibnamefont{et~al.}, \bibinfo{journal}{Phys. Rev. Lett.}
  \textbf{\bibinfo{volume}{126}}, \bibinfo{pages}{076602}
  (\bibinfo{year}{2021}).

\bibitem[{\citenamefont{Jo et~al.}(2020)\citenamefont{Jo, Kuthanazhi, Wu,
  Timmons, Kim, Zhou, Wang, Ueland, Palasyuk, Ryan et~al.}}]{Jo20}
\bibinfo{author}{\bibfnamefont{N.~H.} \bibnamefont{Jo}},
  \bibinfo{author}{\bibfnamefont{B.}~\bibnamefont{Kuthanazhi}},
  \bibinfo{author}{\bibfnamefont{Y.}~\bibnamefont{Wu}},
  \bibinfo{author}{\bibfnamefont{E.}~\bibnamefont{Timmons}},
  \bibinfo{author}{\bibfnamefont{T.-H.} \bibnamefont{Kim}},
  \bibinfo{author}{\bibfnamefont{L.}~\bibnamefont{Zhou}},
  \bibinfo{author}{\bibfnamefont{L.-L.} \bibnamefont{Wang}},
  \bibinfo{author}{\bibfnamefont{B.~G.} \bibnamefont{Ueland}},
  \bibinfo{author}{\bibfnamefont{A.}~\bibnamefont{Palasyuk}},
  \bibinfo{author}{\bibfnamefont{D.~H.} \bibnamefont{Ryan}},
  \bibnamefont{et~al.}, \bibinfo{journal}{Phys. Rev. B}
  \textbf{\bibinfo{volume}{101}}, \bibinfo{pages}{140402}
  (\bibinfo{year}{2020}).

\bibitem[{\citenamefont{Sanjeewa et~al.}(2020)\citenamefont{Sanjeewa, Xing,
  Taddei, Parker, Custelcean, dela Cruz, and Sefat}}]{Sanjeewa20}
\bibinfo{author}{\bibfnamefont{L.~D.} \bibnamefont{Sanjeewa}},
  \bibinfo{author}{\bibfnamefont{J.}~\bibnamefont{Xing}},
  \bibinfo{author}{\bibfnamefont{K.~M.} \bibnamefont{Taddei}},
  \bibinfo{author}{\bibfnamefont{D.}~\bibnamefont{Parker}},
  \bibinfo{author}{\bibfnamefont{R.}~\bibnamefont{Custelcean}},
  \bibinfo{author}{\bibfnamefont{C.}~\bibnamefont{dela Cruz}},
  \bibnamefont{and} \bibinfo{author}{\bibfnamefont{A.~S.} \bibnamefont{Sefat}},
  \bibinfo{journal}{Phys. Rev. B} \textbf{\bibinfo{volume}{102}},
  \bibinfo{pages}{104404} (\bibinfo{year}{2020}).

\bibitem[{\citenamefont{Taddei et~al.}(2020)\citenamefont{Taddei, Lin,
  Sanjeewa, Xing, dela Cruz, Sefat, and Parker}}]{Taddei20}
\bibinfo{author}{\bibfnamefont{K.}~\bibnamefont{Taddei}},
  \bibinfo{author}{\bibfnamefont{L.}~\bibnamefont{Lin}},
  \bibinfo{author}{\bibfnamefont{L.}~\bibnamefont{Sanjeewa}},
  \bibinfo{author}{\bibfnamefont{J.}~\bibnamefont{Xing}},
  \bibinfo{author}{\bibfnamefont{C.}~\bibnamefont{dela Cruz}},
  \bibinfo{author}{\bibfnamefont{A.}~\bibnamefont{Sefat}}, \bibnamefont{and}
  \bibinfo{author}{\bibfnamefont{D.}~\bibnamefont{Parker}},
  \bibinfo{journal}{arXiv: 2012.01555}  (\bibinfo{year}{2020}).

\bibitem[{\citenamefont{Canfield et~al.}(2016)\citenamefont{Canfield, Kong,
  Kaluarachchi, and Jo}}]{Canfield16}
\bibinfo{author}{\bibfnamefont{P.~C.} \bibnamefont{Canfield}},
  \bibinfo{author}{\bibfnamefont{T.}~\bibnamefont{Kong}},
  \bibinfo{author}{\bibfnamefont{U.~S.} \bibnamefont{Kaluarachchi}},
  \bibnamefont{and} \bibinfo{author}{\bibfnamefont{N.~H.} \bibnamefont{Jo}},
  \bibinfo{journal}{Philosophical Magazine} \textbf{\bibinfo{volume}{96}},
  \bibinfo{pages}{84} (\bibinfo{year}{2016}).

\bibitem[{Can()}]{CanfieldCCS}
\urlprefix\url{https://lspceramics.com/canfield-crucible-sets-2/}.

\bibitem[{\citenamefont{Canfield}(2019)}]{Canfield19}
\bibinfo{author}{\bibfnamefont{P.~C.} \bibnamefont{Canfield}},
  \bibinfo{journal}{Reports on Progress in Physics}
  \textbf{\bibinfo{volume}{83}}, \bibinfo{pages}{016501}
  (\bibinfo{year}{2019}).

\bibitem[{\citenamefont{Gati et~al.}(2019)\citenamefont{Gati, Drachuck, Xiang,
  Wang, Bud'ko, and Canfield}}]{Gati19}
\bibinfo{author}{\bibfnamefont{E.}~\bibnamefont{Gati}},
  \bibinfo{author}{\bibfnamefont{G.}~\bibnamefont{Drachuck}},
  \bibinfo{author}{\bibfnamefont{L.}~\bibnamefont{Xiang}},
  \bibinfo{author}{\bibfnamefont{L.-L.} \bibnamefont{Wang}},
  \bibinfo{author}{\bibfnamefont{S.~L.} \bibnamefont{Bud'ko}},
  \bibnamefont{and} \bibinfo{author}{\bibfnamefont{P.~C.}
  \bibnamefont{Canfield}}, \bibinfo{journal}{Rev. Sci. Instrum.}
  \textbf{\bibinfo{volume}{90}}, \bibinfo{pages}{023911}
  (\bibinfo{year}{2019}).

\bibitem[{\citenamefont{Bud'ko et~al.}(1984)\citenamefont{Bud'ko, Voronovskii,
  Gapotchenko, and ltskevich}}]{Budko84}
\bibinfo{author}{\bibfnamefont{S.~L.} \bibnamefont{Bud'ko}},
  \bibinfo{author}{\bibfnamefont{A.~N.} \bibnamefont{Voronovskii}},
  \bibinfo{author}{\bibfnamefont{A.~G.} \bibnamefont{Gapotchenko}},
  \bibnamefont{and} \bibinfo{author}{\bibfnamefont{E.~S.}
  \bibnamefont{ltskevich}}, \bibinfo{journal}{Zh. Eksp. Teor. Fiz.}
  \textbf{\bibinfo{volume}{86}}, \bibinfo{pages}{778} (\bibinfo{year}{1984}).

\bibitem[{\citenamefont{Torikachvili et~al.}(2015)\citenamefont{Torikachvili,
  Kim, Colombier, Bud'ko, and Canfield}}]{Torikachvili15}
\bibinfo{author}{\bibfnamefont{M.~S.} \bibnamefont{Torikachvili}},
  \bibinfo{author}{\bibfnamefont{S.~K.} \bibnamefont{Kim}},
  \bibinfo{author}{\bibfnamefont{E.}~\bibnamefont{Colombier}},
  \bibinfo{author}{\bibfnamefont{S.~L.} \bibnamefont{Bud'ko}},
  \bibnamefont{and} \bibinfo{author}{\bibfnamefont{P.~C.}
  \bibnamefont{Canfield}}, \bibinfo{journal}{Rev. Sci. Instrum.}
  \textbf{\bibinfo{volume}{86}}, \bibinfo{pages}{123904}
  (\bibinfo{year}{2015}).

\bibitem[{\citenamefont{Eiling and Schilling}(1981)}]{Eiling81}
\bibinfo{author}{\bibfnamefont{A.}~\bibnamefont{Eiling}} \bibnamefont{and}
  \bibinfo{author}{\bibfnamefont{J.~S.} \bibnamefont{Schilling}},
  \bibinfo{journal}{Journal of Physics F: Metal Physics}
  \textbf{\bibinfo{volume}{11}}, \bibinfo{pages}{623} (\bibinfo{year}{1981}).

\bibitem[{\citenamefont{Khasanov et~al.}(2016)\citenamefont{Khasanov, Guguchia,
  Maisuradze, Andreica, Elender, Raselli, Shermadini, Goko, Knecht, Morenzoni
  et~al.}}]{Khasanov16}
\bibinfo{author}{\bibfnamefont{R.}~\bibnamefont{Khasanov}},
  \bibinfo{author}{\bibfnamefont{Z.}~\bibnamefont{Guguchia}},
  \bibinfo{author}{\bibfnamefont{A.}~\bibnamefont{Maisuradze}},
  \bibinfo{author}{\bibfnamefont{D.}~\bibnamefont{Andreica}},
  \bibinfo{author}{\bibfnamefont{M.}~\bibnamefont{Elender}},
  \bibinfo{author}{\bibfnamefont{A.}~\bibnamefont{Raselli}},
  \bibinfo{author}{\bibfnamefont{Z.}~\bibnamefont{Shermadini}},
  \bibinfo{author}{\bibfnamefont{T.}~\bibnamefont{Goko}},
  \bibinfo{author}{\bibfnamefont{F.}~\bibnamefont{Knecht}},
  \bibinfo{author}{\bibfnamefont{E.}~\bibnamefont{Morenzoni}},
  \bibnamefont{et~al.}, \bibinfo{journal}{High Pressure Research}
  \textbf{\bibinfo{volume}{36}}, \bibinfo{pages}{140} (\bibinfo{year}{2016}).

\bibitem[{\citenamefont{Smith and Chu}(1967)}]{Smith67}
\bibinfo{author}{\bibfnamefont{T.~F.} \bibnamefont{Smith}} \bibnamefont{and}
  \bibinfo{author}{\bibfnamefont{C.~W.} \bibnamefont{Chu}},
  \bibinfo{journal}{Phys. Rev.} \textbf{\bibinfo{volume}{159}},
  \bibinfo{pages}{353} (\bibinfo{year}{1967}).

\bibitem[{\citenamefont{Hohenberg and Kohn}(1964)}]{Hohenberg64}
\bibinfo{author}{\bibfnamefont{P.}~\bibnamefont{Hohenberg}} \bibnamefont{and}
  \bibinfo{author}{\bibfnamefont{W.}~\bibnamefont{Kohn}},
  \bibinfo{journal}{Phys. Rev.} \textbf{\bibinfo{volume}{136}},
  \bibinfo{pages}{B864} (\bibinfo{year}{1964}).

\bibitem[{\citenamefont{Kohn and Sham}(1965)}]{Kohn65}
\bibinfo{author}{\bibfnamefont{W.}~\bibnamefont{Kohn}} \bibnamefont{and}
  \bibinfo{author}{\bibfnamefont{L.~J.} \bibnamefont{Sham}},
  \bibinfo{journal}{Phys. Rev.} \textbf{\bibinfo{volume}{140}},
  \bibinfo{pages}{A1133} (\bibinfo{year}{1965}).

\bibitem[{\citenamefont{Perdew et~al.}(1996)\citenamefont{Perdew, Burke, and
  Ernzerhof}}]{Perdew96}
\bibinfo{author}{\bibfnamefont{J.~P.} \bibnamefont{Perdew}},
  \bibinfo{author}{\bibfnamefont{K.}~\bibnamefont{Burke}}, \bibnamefont{and}
  \bibinfo{author}{\bibfnamefont{M.}~\bibnamefont{Ernzerhof}},
  \bibinfo{journal}{Phys. Rev. Lett.} \textbf{\bibinfo{volume}{77}},
  \bibinfo{pages}{3865} (\bibinfo{year}{1996}).

\bibitem[{\citenamefont{Bl\"ochl}(1994)}]{Bloechl94}
\bibinfo{author}{\bibfnamefont{P.~E.} \bibnamefont{Bl\"ochl}},
  \bibinfo{journal}{Phys. Rev. B} \textbf{\bibinfo{volume}{50}},
  \bibinfo{pages}{17953} (\bibinfo{year}{1994}).

\bibitem[{\citenamefont{Kresse and Joubert}(1999)}]{Boechl99}
\bibinfo{author}{\bibfnamefont{G.}~\bibnamefont{Kresse}} \bibnamefont{and}
  \bibinfo{author}{\bibfnamefont{D.}~\bibnamefont{Joubert}},
  \bibinfo{journal}{Phys. Rev. B} \textbf{\bibinfo{volume}{59}},
  \bibinfo{pages}{1758} (\bibinfo{year}{1999}).

\bibitem[{\citenamefont{Kresse and Hafner}(1993)}]{Kresse93}
\bibinfo{author}{\bibfnamefont{G.}~\bibnamefont{Kresse}} \bibnamefont{and}
  \bibinfo{author}{\bibfnamefont{J.}~\bibnamefont{Hafner}},
  \bibinfo{journal}{Phys. Rev. B} \textbf{\bibinfo{volume}{47}},
  \bibinfo{pages}{558} (\bibinfo{year}{1993}).

\bibitem[{\citenamefont{Kresse and Furthm\"uller}(1996)}]{Kresse96}
\bibinfo{author}{\bibfnamefont{G.}~\bibnamefont{Kresse}} \bibnamefont{and}
  \bibinfo{author}{\bibfnamefont{J.}~\bibnamefont{Furthm\"uller}},
  \bibinfo{journal}{Phys. Rev. B} \textbf{\bibinfo{volume}{54}},
  \bibinfo{pages}{11169} (\bibinfo{year}{1996}).

\bibitem[{\citenamefont{Kresse and Furthmüller}(1996)}]{Kresse9615}
\bibinfo{author}{\bibfnamefont{G.}~\bibnamefont{Kresse}} \bibnamefont{and}
  \bibinfo{author}{\bibfnamefont{J.}~\bibnamefont{Furthmüller}},
  \bibinfo{journal}{Computational Materials Science}
  \textbf{\bibinfo{volume}{6}}, \bibinfo{pages}{15} (\bibinfo{year}{1996}).

\bibitem[{\citenamefont{Monkhorst and Pack}(1976)}]{Monkhorst76}
\bibinfo{author}{\bibfnamefont{H.~J.} \bibnamefont{Monkhorst}}
  \bibnamefont{and} \bibinfo{author}{\bibfnamefont{J.~D.} \bibnamefont{Pack}},
  \bibinfo{journal}{Phys. Rev. B} \textbf{\bibinfo{volume}{13}},
  \bibinfo{pages}{5188} (\bibinfo{year}{1976}).

\bibitem[{\citenamefont{Fisher and Langer}(1968)}]{Fisher68}
\bibinfo{author}{\bibfnamefont{M.~E.} \bibnamefont{Fisher}} \bibnamefont{and}
  \bibinfo{author}{\bibfnamefont{J.~S.} \bibnamefont{Langer}},
  \bibinfo{journal}{Phys. Rev. Lett.} \textbf{\bibinfo{volume}{20}},
  \bibinfo{pages}{665} (\bibinfo{year}{1968}).

\bibitem[{\citenamefont{Taufour et~al.}(2016)\citenamefont{Taufour,
  Kaluarachchi, Khasanov, Nguyen, Guguchia, Biswas, Bonf\`a, De~Renzi, Lin, Kim
  et~al.}}]{Taufour16}
\bibinfo{author}{\bibfnamefont{V.}~\bibnamefont{Taufour}},
  \bibinfo{author}{\bibfnamefont{U.~S.} \bibnamefont{Kaluarachchi}},
  \bibinfo{author}{\bibfnamefont{R.}~\bibnamefont{Khasanov}},
  \bibinfo{author}{\bibfnamefont{M.~C.} \bibnamefont{Nguyen}},
  \bibinfo{author}{\bibfnamefont{Z.}~\bibnamefont{Guguchia}},
  \bibinfo{author}{\bibfnamefont{P.~K.} \bibnamefont{Biswas}},
  \bibinfo{author}{\bibfnamefont{P.}~\bibnamefont{Bonf\`a}},
  \bibinfo{author}{\bibfnamefont{R.}~\bibnamefont{De~Renzi}},
  \bibinfo{author}{\bibfnamefont{X.}~\bibnamefont{Lin}},
  \bibinfo{author}{\bibfnamefont{S.~K.} \bibnamefont{Kim}},
  \bibnamefont{et~al.}, \bibinfo{journal}{Phys. Rev. Lett.}
  \textbf{\bibinfo{volume}{117}}, \bibinfo{pages}{037207}
  (\bibinfo{year}{2016}).

\bibitem[{\citenamefont{Gati et~al.}(2021)\citenamefont{Gati, Wilde, Khasanov,
  Xiang, Dissanayake, Gupta, Matsuda, Ye, Haberl, Kaluarachchi
  et~al.}}]{Gati21}
\bibinfo{author}{\bibfnamefont{E.}~\bibnamefont{Gati}},
  \bibinfo{author}{\bibfnamefont{J.~M.} \bibnamefont{Wilde}},
  \bibinfo{author}{\bibfnamefont{R.}~\bibnamefont{Khasanov}},
  \bibinfo{author}{\bibfnamefont{L.}~\bibnamefont{Xiang}},
  \bibinfo{author}{\bibfnamefont{S.}~\bibnamefont{Dissanayake}},
  \bibinfo{author}{\bibfnamefont{R.}~\bibnamefont{Gupta}},
  \bibinfo{author}{\bibfnamefont{M.}~\bibnamefont{Matsuda}},
  \bibinfo{author}{\bibfnamefont{F.}~\bibnamefont{Ye}},
  \bibinfo{author}{\bibfnamefont{B.}~\bibnamefont{Haberl}},
  \bibinfo{author}{\bibfnamefont{U.}~\bibnamefont{Kaluarachchi}},
  \bibnamefont{et~al.}, \bibinfo{journal}{Phys. Rev. B}
  \textbf{\bibinfo{volume}{103}}, \bibinfo{pages}{075111}
  (\bibinfo{year}{2021}).

\bibitem[{\citenamefont{Khasanov et~al.}(2017)\citenamefont{Khasanov, Amato,
  Bonf{\`{a}}, Guguchia, Luetkens, Morenzoni, Renzi, and
  Zhigadlo}}]{Khasanov17}
\bibinfo{author}{\bibfnamefont{R.}~\bibnamefont{Khasanov}},
  \bibinfo{author}{\bibfnamefont{A.}~\bibnamefont{Amato}},
  \bibinfo{author}{\bibfnamefont{P.}~\bibnamefont{Bonf{\`{a}}}},
  \bibinfo{author}{\bibfnamefont{Z.}~\bibnamefont{Guguchia}},
  \bibinfo{author}{\bibfnamefont{H.}~\bibnamefont{Luetkens}},
  \bibinfo{author}{\bibfnamefont{E.}~\bibnamefont{Morenzoni}},
  \bibinfo{author}{\bibfnamefont{R.~D.} \bibnamefont{Renzi}}, \bibnamefont{and}
  \bibinfo{author}{\bibfnamefont{N.~D.} \bibnamefont{Zhigadlo}},
  \bibinfo{journal}{Journal of Physics: Condensed Matter}
  \textbf{\bibinfo{volume}{29}}, \bibinfo{pages}{164003}
  (\bibinfo{year}{2017}).

\bibitem[{\citenamefont{Birch}(1947)}]{Birch47}
\bibinfo{author}{\bibfnamefont{F.}~\bibnamefont{Birch}},
  \bibinfo{journal}{Phys. Rev.} \textbf{\bibinfo{volume}{71}},
  \bibinfo{pages}{809} (\bibinfo{year}{1947}).

\bibitem[{\citenamefont{Murnaghan}(1944)}]{Murnaghan44}
\bibinfo{author}{\bibfnamefont{F.~D.} \bibnamefont{Murnaghan}},
  \bibinfo{journal}{Proceedings of the National Academy of Sciences}
  \textbf{\bibinfo{volume}{30}}, \bibinfo{pages}{244} (\bibinfo{year}{1944}).

\bibitem[{\citenamefont{Harl et~al.}(2010)\citenamefont{Harl, Schimka, and
  Kresse}}]{Harl10}
\bibinfo{author}{\bibfnamefont{J.}~\bibnamefont{Harl}},
  \bibinfo{author}{\bibfnamefont{L.}~\bibnamefont{Schimka}}, \bibnamefont{and}
  \bibinfo{author}{\bibfnamefont{G.}~\bibnamefont{Kresse}},
  \bibinfo{journal}{Phys. Rev. B} \textbf{\bibinfo{volume}{81}},
  \bibinfo{pages}{115126} (\bibinfo{year}{2010}).

\bibitem[{\citenamefont{Shermadini et~al.}(2017)\citenamefont{Shermadini,
  Khasanov, Elender, Simutis, Guguchia, Kamenev, and Amato}}]{Shermandini17}
\bibinfo{author}{\bibfnamefont{Z.}~\bibnamefont{Shermadini}},
  \bibinfo{author}{\bibfnamefont{R.}~\bibnamefont{Khasanov}},
  \bibinfo{author}{\bibfnamefont{M.}~\bibnamefont{Elender}},
  \bibinfo{author}{\bibfnamefont{G.}~\bibnamefont{Simutis}},
  \bibinfo{author}{\bibfnamefont{Z.}~\bibnamefont{Guguchia}},
  \bibinfo{author}{\bibfnamefont{K.}~\bibnamefont{Kamenev}}, \bibnamefont{and}
  \bibinfo{author}{\bibfnamefont{A.}~\bibnamefont{Amato}},
  \bibinfo{journal}{High Pressure Research} \textbf{\bibinfo{volume}{37}},
  \bibinfo{pages}{449} (\bibinfo{year}{2017}).

\end{thebibliography}

\end{document}